%% file: bare_conf.tex
\documentclass[conference]{IEEEtran}
\ifCLASSINFOpdf
\else
\fi
\usepackage{amsmath}
\usepackage{graphicx}
\usepackage{booktabs}
\usepackage{makecell}
\usepackage{pifont}
\usepackage{xcolor}
\usepackage{cite}
\usepackage{bm}
\usepackage{float}
\usepackage{amssymb,amsfonts}
\usepackage{mathtools, nccmath}
\usepackage{textcomp}
\usepackage[font=small]{caption}
\usepackage{subcaption}
\usepackage{nicematrix}
\usepackage[shortlabels,inline]{enumitem}
\usepackage{comment}
\usepackage{cuted}
\usepackage[para,flushleft]{threeparttable}
\usepackage{adjustbox}
\usepackage{color, colortbl}
\usepackage{soul}
\usepackage{multirow}
\usepackage{tikz}
\usepackage{pgfplots}
\usepackage{algorithm}
\usepackage{siunitx}
\usepackage[noend]{algpseudocode}%
\usepackage[hidelinks]{hyperref}

\newcommand{\cmark}{\textcolor{green}{\ding{51}}}%
\newcommand{\xmark}{\textcolor{red}{\ding{55}}}%

\makeatletter
\newcommand{\linebreakand}{%
  \end{@IEEEauthorhalign}
  \hfill\mbox{}\par
  \mbox{}\hfill\begin{@IEEEauthorhalign}
}
\makeatother

\begin{document}
\title{SMOF: Streaming Modern CNNs on FPGAs with Smart Off-Chip Eviction}

\author{
    \IEEEauthorblockN{Petros Toupas\IEEEauthorrefmark{1}$^{1,2}$}
    \and
    \IEEEauthorblockN{Zhewen Yu\IEEEauthorrefmark{1}$^{1}$}
    \and
    \IEEEauthorblockN{Christos-Savvas Bouganis$^1$}
    \and
    \IEEEauthorblockN{Dimitrios Tzovaras$^2$}
    \linebreakand
    \IEEEauthorblockA{
    $^1$
    Imperial College London\\
    \{p.toupas21, zhewen.yu18, \\christos-savvas.bouganis\}@imperial.ac.uk}
    \and
    \IEEEauthorblockA{
    $^2$
    Information Technologies Institute\\
    Centre of Research and Technology Hellas\\
    \{ptoupas,dimitrios.tzovaras\}@iti.gr}
}

\maketitle

\begingroup\renewcommand\thefootnote{\IEEEauthorrefmark{1}}
\footnotetext{\textit{Equal contribution}}
\endgroup

\begin{abstract}
Convolutional Neural Networks (CNNs) have demonstrated their effectiveness in numerous vision tasks. However, their high processing requirements necessitate efficient hardware acceleration to meet the application’s performance targets. In the space of FPGAs, streaming-based dataflow architectures are often adopted by users, as significant performance gains can be achieved through layer-wise pipelining and reduced off-chip memory access by retaining data on-chip. However, modern topologies, such as the UNet, YOLO, or X3D models, utilise long skip connections, requiring significant on-chip storage and thus limiting the performance achieved by such system architectures. The paper addresses the above limitation by introducing weight and activation eviction mechanisms to off-chip memory along the computational pipeline, taking into account the available compute and memory resources. The proposed mechanism is incorporated into an existing toolflow, expanding the design space by utilising off-chip memory as a buffer. This enables the mapping of such modern CNNs to devices with limited on-chip memory, under the streaming architecture design approach. SMOF has demonstrated the capacity to deliver competitive and, in some cases, state-of-the-art performance across a spectrum of computer vision tasks, achieving up to 10.65$\times$ throughput improvement compared to previous works. The tool is available at \underline{\url{https://github.com/ICIdsl/smof.git}.}
\end{abstract}

\IEEEpeerreviewmaketitle

\input{Introduction/introduction.tex}

\input{RelatedWork/relatedwork.tex}

\input{Methodology/methodology.tex}

\input{Implementation/implementation.tex}

\input{Evaluation/evaluation.tex}

\input{Conclusion/conclusion.tex}

\section*{Acknowledgment}
For the purpose of open access, the authors have applied a Creative Commons Attribution (CC BY) license to any Accepted Manuscript version arising.

\bibliographystyle{IEEEtran}
\bibliography{references}

\end{document}

%% file: Introduction/introduction.tex
\section{Introduction} \label{introduction}
Convolutional Neural Networks (CNNs) have been applied to a wide range of applications, and they are especially successful in dealing with vision tasks such as image classification, object detection, and semantic segmentation. Due to their computational intensity, a large amount of work has been done to accelerate CNNs in hardware, with Field-Programmable Gate Arrays (FPGA) devices serving as a high-performing and energy-efficient platform of choice.

In the realm of mapping CNNs to FPGAs, there are two alternatives. The first one involves building a unified accelerator architecture, embracing the concept of ``one-fits-all" to minimise the engineering effort of re-development \cite{gokhale2017snowflake, guo2017angel}. The second approach entails building a streaming dataflow architecture, by customising a layerwise pipeline for each specific workload. In this work, we focus on the design of the streaming architecture which is often the preferred choice under a fixed model workload.

To facilitate the end-to-end deployment of streaming architectures, toolflows such as fpgaConvNet\cite{Venieris2019}, FINN \cite{umuroglu2017finn}, hls4ml \cite{fahim2021hls4ml} and HPIPE \cite{hall2020tensorflow} have been developed, that generate HLS or RTL designs from high-level network descriptions (Tensorflow, PyTorch, ONNX). In terms of memory storage, the common practice employed in these toolflows involves confining most data movements within the FPGA chip \cite{umuroglu2017finn, Venieris2019}. This practice entails storing weights statically on-chip in a read-only format, while activations are transferred between layers employing FIFOs with handshake interfaces. However, as networks expand to accommodate millions of parameters and complex skip connections, scaling this approach becomes a challenging task.
For example, the UNet model \cite{ronneberger2015u}, a popular benchmark utilised in semantic segmentation, has a significant operation count at 130.1G. Moreover, UNet contains lengthy hierarchical branch connections, which require the extensive placement of FIFOs to synchronise data streams across its branching structure. Similar model characteristics and specifics are encountered in YOLO \cite{redmon2016you}, X3D \cite{Feichtenhofer2020} and many other models across various domains.

In fact, existing streaming architectures primarily concentrate on the optimisation of computation engines and on-chip memory resources. Off-chip memory is mainly utilised for transferring data at the input and output pipeline stages, prohibiting the utilisation of existing toolflows when on-chip memory requirements exceed the available resources.
As such, existing streaming architectures are often limited by the on-chip memory, while leaving off-chip memory underutilised. 

In this paper, we introduce a groundbreaking memory optimisation methodology that systematically considers the allocation and utilisation of both on-chip and off-chip memory within a layerwise pipelined, streaming architecture. Our contributions extend beyond conventional approaches, aiming to address the limitations encountered in scaling networks with substantial numbers of parameters and intricate connections that require excessive buffering. Our key contributions are:
\begin{itemize}
    \item We propose the first streaming CNN accelerator that can partially offload weights and activations to the off-chip memory, without stalling the computation pipeline.
    \item We introduce a subgraph-based partitioning methodology offering the latency-throughput design trade-off, by exploiting the device reconfigurability of FPGAs.
    \item We propose a Design Space Exploration (DSE) methodology that relies on a greedy and iterative optimisation algorithm to automatically explore and determine the optimal memory and partitioning configuration.
    \item We accelerate a wide range of CNN benchmarks on a diverse spectrum of computer vision tasks using the proposed methodology, demonstrating competitive and even state-of-the-art performance, particularly on networks with complex, hierarchical skip connections. 
\end{itemize}

%% file: RelatedWork/relatedwork.tex
\begin{table*}[ht]
\centering
\caption{Comparison of our work with related works.}
\label{tab:related_works}
\begin{tabular}{lcccccc|ccccccc}
\toprule
Works & \makecell[c]{Snowflake \\ \cite{gokhale2017snowflake}} & \makecell[c]{Angle-eye \\ \cite{guo2017angel}} & \makecell[c]{Brainwave \\ \cite{fowers2018configurable}}  & \cite{shen2018towards} & \cite{liu2018optimizing} & \makecell[c]{Vitis AI \\ \cite{kathail2020xilinx}} & \makecell[c]{DeepBurning \\ \cite{wang2016deepburning}} & \makecell[c]{FINN \\ \cite{umuroglu2017finn}} & \makecell[c]{DeepBuilder\\ \cite{zhang2018dnnbuilder}} & \makecell[c]{HPIPE \\ \cite{hall2020tensorflow}} & \makecell[c]{SMOF \\ Our work} \\
\midrule
Architecture Style & \multicolumn{6}{c|}{Single Engine Architectures} & \multicolumn{5}{c}{Streaming Architectures} \\
\midrule
Classification & \cmark & \cmark & \cmark & \cmark & \xmark & \cmark & \cmark & \cmark &\cmark & \cmark & \cmark \\
Detection & \xmark & \cmark & \xmark & \xmark & \xmark & \cmark & \xmark & \xmark &\cmark & \cmark & \cmark \\
Segmentation & \xmark & \xmark & \xmark & \xmark & \cmark & \cmark & \xmark & \xmark & \xmark & \xmark & \cmark \\
\midrule
2D CNN & \cmark & \cmark & \cmark & \cmark & \cmark & \cmark & \cmark & \cmark & \cmark & \cmark & \cmark \\
3D CNN & \xmark & \xmark & \xmark & \cmark & \xmark & \xmark & \xmark & \xmark & \xmark & \xmark & \cmark \\
\bottomrule
\end{tabular}
\end{table*}

\section{Related Work} \label{relatedwork}

\subsection{Streaming Architectures}
Various approaches have been proposed for building efficient neural network accelerators, such as systolic arrays \cite{wei2017automated, samajdar2018scale}, vector processors \cite{gokhale2017snowflake, guo2017angel} and streaming architectures \cite{wang2016deepburning, umuroglu2017finn}. Among these approaches, the streaming architectures achieve state-of-the-art performance when the targeted model is fixed due to their layer-wise pipelining architecture and model-specific hardware customisations. However, due to their design complexity and the variability of modern CNNs, it is not feasible to implement streaming architectures manually. Therefore, automated toolflows have been proposed in previous literature to generate efficient streaming architectures. DeepBurning \cite{wang2016deepburning}, fpgaConvNet \cite{Venieris2019}, FINN \cite{umuroglu2017finn} and HPIPE \cite{hall2020tensorflow} are well-known examples in this line of reasearch.

Currently, the emphasis in these toolflows has been placed into the optimisation of computation engines. FINN \cite{umuroglu2017finn} focuses on the computation of Binarized Neural Networks (BNN) by optimising BNN-specific operations such as XNOR and Popcount, resulting in ultra-low resource utilisation compared with normal fixed-point operations. DeepBurning \cite{luo2023deepburning} investigated DSP packing to support mixed-precision quantization, where the heterogeneous computation engines operate on different precisions per-layer. Recent work on HPIPE \cite{hall2020tensorflow} and fpgaConvNet \cite{montgomerie2023pass} exploit the data sparsity to skip any zero multiplication for performance gains. The skipping is controlled by either compile-time or run-time scheduling.

In a typical streaming architecture, most memory footprints are on-chip, as all the weights are stored in BRAMs and the activations are buffered in FIFOs. To efficiently squeeze on-chip memory resources, FINN \cite{petrica2020memory} considered the overclocking of BRAMs, which provides virtual memory ports to mitigate the mismatches between desired and available
memory shapes, and eventually improve the utilisation efficiency of BRAMs. HPIPE \cite{hall2020tensorflow} compressed the sparse weight parameters to the run-length encoded format, reducing on-chip storage with a small decoding overhead during computation. To overcome on-chip memory limitations, recent works \cite{alonso2021elastic, Ibrahim2023extending} consider partitioning the computational pipeline across multiple FPGA devices, in the expense of under-utilising sometimes the individual devices.

While single engine architectures utilise tiling to overcome bottlenecks imposed by on-chip storage constraints, streaming architectures often struggle to overcome such limitations, especially on networks with extensive parameters and intricate connections that require significant buffering. Our proposed solution, extended certain concepts introduced in \cite{montgomerie2023satay, yu2023autows}, tackles these challenges by strategically managing the allocation and utilisation of both on-chip and off-chip memory by partially offloading weights and activations to the off-chip memory, without stalling the computation pipeline of a streaming architecture.

\subsection{Accelerate Vision Tasks}
In terms of supported CNN workloads, many accelerator architectures focus on the image classification task (TABLE~\ref{tab:related_works}). These architectures have been able to accommodate popular networks such as VGG \cite{simonyan2014very}, GoogLeNet \cite{szegedy2015going}, ResNet \cite{he2016deep} and MobileNet \cite{howard2017mobilenets} on the datasets such as MNIST \cite{lecun1998mnist}, CIFAR \cite{krizhevsky2009learning} and ImageNet \cite{deng2009imagenet}. However, despite the substantial emphasis on image classification, there are several other crucial vision tasks in the realm of computer vision, including but not limited to object detection, semantic segmentation, as well as video and volumetric data analysis. Compared with image classification, these tasks usually demand higher computational resources and more complex network architectures, creating new challenges for hardware accelerator architectures.

Angle-eye \cite{guo2017angel} supported the deployment of the well-known YOLO model \cite{redmon2016you} for the object detection task, where the CNN backbone is quantized to 8 bits and computed on an FPGA device using a reconfigurable vector processor, while the pre-process and post-process computation is offloaded to a CPU. A similar approach was adopted by Nakahara \textit{et al.} \cite{nakahara2018lightweight}, but the authors chose to binarize the YOLO model for an efficient hardware implementation. HPIPE \cite{anupreetham2023high} and SATAY \cite{montgomerie2023satay} are two recent works that consider the acceleration of object detection workloads using the streaming architecture, where the main challenges addressed are the design of high performance Non-Maximum Suppression and HardSwish modules.

The UNet model \cite{ronneberger2015u} is a popular benchmark utilised in the semantic segmentation task, and the model contains both downsampling and upsampling operations in order to generate accurate masks. As the acceleration of convolutional layer has been well investigated, the main consideration when designing hardware accelerators for this load is the investigation of deconvolution algorithms for efficient upsampling. Liu \textit{et al.} \cite{liu2018optimizing} designed parameterised convolution and deconvolution engines with shared input buffers. However, there has not been any attempt to deploy UNet under a streaming architecture, as the long skip connection between downsampling and upsampling layers requires deep on-chip buffers.

To efficiently interpret information across video frames, 3D CNNs are often favoured with an additional dimension inside the convolution kernel. Shen \textit{et al.} \cite{shen2018towards} developed a template-based architecture targeting both 2D and 3D CNN models, utilising the Winograd algorithm to efficiently reduce the number of multiplications in convolutions and hence the computational complexity overall. Additionally, they introduced an analytical method for efficiently exploring the design space to identify optimal tiling strategies. fpgaHART \cite{toupas2023fpgahart} evaluated a variety of 3D CNN models using streaming accelerator architectures. However, it could not target networks like 3D UNet due to the large on-chip buffer overhead from the long skip connection. HARFLOW3D \cite{toupas2023harflow3d} presented a latency-driven streaming accelerator toolflow targeting modern 3D CNN models. The authors introduced parameterized hardware building blocks that support runtime parameterization in a time-shared manner, but with limited supported layer depth.

Our approach, because of the combination of the streaming architecture and the proposed smart off-chip data eviction, is able to support and deliver high throughput designs on a diverse spectrum of computer vision tasks using state-of-the-art 2D and 3D CNN models, as shown in TABLE~\ref{relatedwork} and \ref{tab:fpga-comparison}.

%% file: Methodology/methodology.tex
\section{Methodology} \label{methodology}
In this section, we propose and formalise a threefold way to mitigate the on-chip memory bottlenecks in streaming architectures, by introducing activation eviction, weight fragmentation and reconfiguration as possible design choices.

\subsection{Activation Eviction}

In streaming design toolflows, the CNN workload is abstracted to a Directed Acyclic Graph (DAG) where vertices represent operations like convolution, pooling, and activation function, while edges represent the data flows between these operations. One critical aspect in streaming architectures is to minimise the occurrence of stalls between the computational operations. In the case of sequential connections between vertices, buffers are inserted to address any instantaneous rate mismatches that might occur between consecutive operations, as the operations may produce or consume a burst of data to sustain their continuous execution. 

Moreover, in instances where the DAG branches into multiple computational paths, the role of buffers becomes even more critical. Branch connections entail diverging paths where different computational branches may operate at varying throughputs or exhibit different processing latencies. Buffers placed at these branching points prevent a fast branch from overwhelming or stalling a slower branch. These buffers help maintain a balanced data flow, enabling synchronization between branches and preventing potential deadlocks.

\begin{figure}[t]
    \centering
    \includegraphics[width=\columnwidth, keepaspectratio]{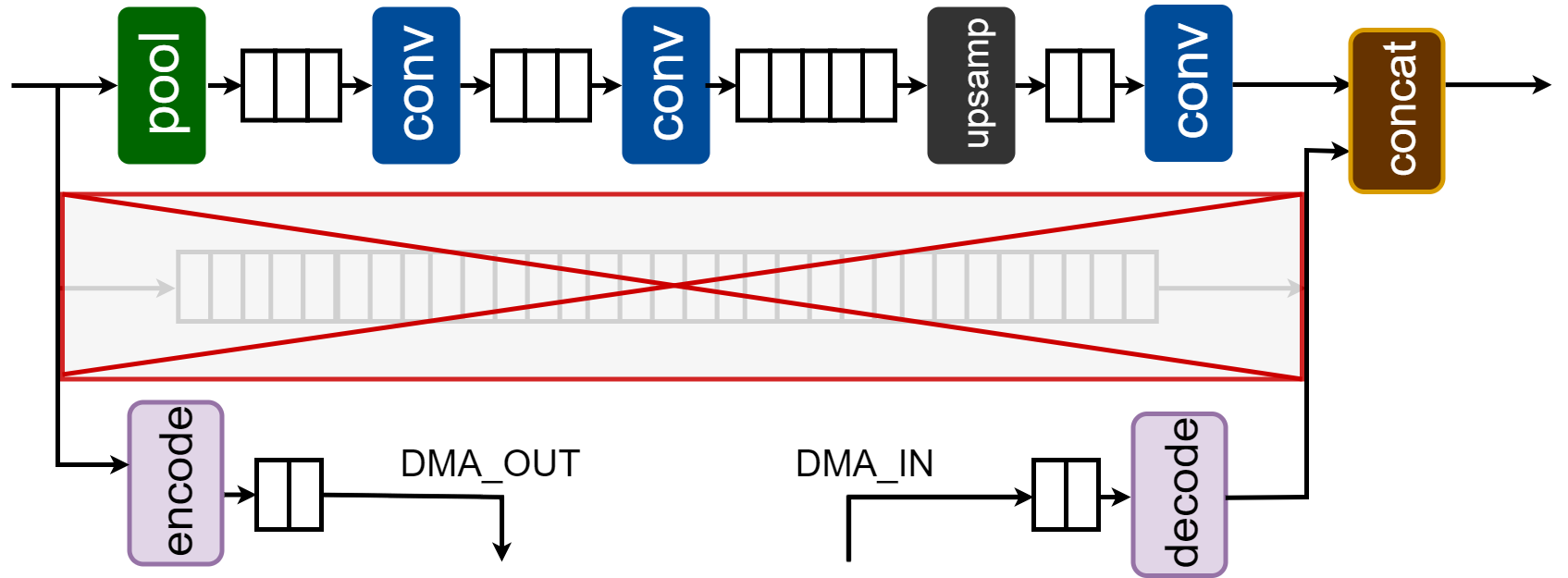}
    \caption{Activation eviction. For the long skip connection, instead of being held at the on-chip buffer, the activation data is pushed to the off-chip memory. We support the lossless encoding schemes, such as RLE and Huffman, to save the off-chip bandwidth.}
    \label{fig:activation_eviction}
\end{figure}

To reduce the on-chip memory resources required by implementing these buffers, we propose the eviction of a subset of activation data to the off-chip memory instead. Specifically, the original deep buffer with the depth of $d$ is replaced with the circuit as Figure~\ref{fig:activation_eviction} shows, which only contains two small FIFOs to only hold the data chunk enough for a DMA burst, and the remaining activation data are pushed to off-chip. At the host side, pointers are used to manage the head and tail positions of the evicted activation data. 

An important question to answer is at which edge should the eviction be applied. A considerable number of design choices are available, as we can choose to evict the edge of any two connected vertices. Let us start with establishing the trade-off between on-chip and off-chip memory resources for transferring activations in the streaming architecture. 

If we denote the total depth of the two small FIFOs after eviction as $d_b'$ and compare it with the original buffer depth of $d_b$, the saving in buffer depth will be: 
\begin{equation}
    \Delta d = d_b - d_b',\;\;s.t.\;\;d_b > max(d_b', t_{db})
    \label{activation buffer depth}
\end{equation}

In this equation, we also apply an extra constraint that the original buffer depth $d_b$ must be larger than the delay of DMA transfers $t_{db}$. Otherwise, the eviction will cause extra stalls in the computational pipeline, downgrading the overall performance.

On the other hand, the extra off-chip bandwidth required, including both writing and  reading, after the eviction can be calculated as,

\begin{equation}
    \Delta BW = r \bar{c} (1+\alpha)
    \label{activation bandwidth}
\end{equation}
where the average data rate is represented as $r$ (words/cycle). Let's use $\bar{c}$  to denote the average compression ratio for DMA transfers, as activations can be encoded losslessly using formats such as Run-Length Encoding (RLE) or Huffman. 

Note that the compression ratio for activation data actually varies on different input images. At design time, we use the average ratio to estimate the bandwidth requirement, and we will analyse the deviation of this estimation and it's performance impact at the end of Section~\ref{Off-Chip Streaming Compression}.

In Equation~\eqref{activation bandwidth}, $\alpha$ is a penalty factor for the reading bandwidth, whose value is always larger than or equal to one. $\alpha$ is only larger than one, if the data order in reading and writing are different, in which case, the off-chip reading becomes random access and the bandwidth is penalised because of that. 

Overall, Equation~\eqref{activation buffer depth} and \eqref{activation bandwidth} establish the trade-off between on-chip and off-chip memory resources for transferring activations in the streaming architecture. In Section~\ref{greedy exploration}, we utilise the ratio of this trade-off to decide where the activations should be evicted.

\subsection{Weight Fragmentation}

\begin{figure}[t]
    \centering
    \includegraphics[width=0.8\columnwidth, keepaspectratio]{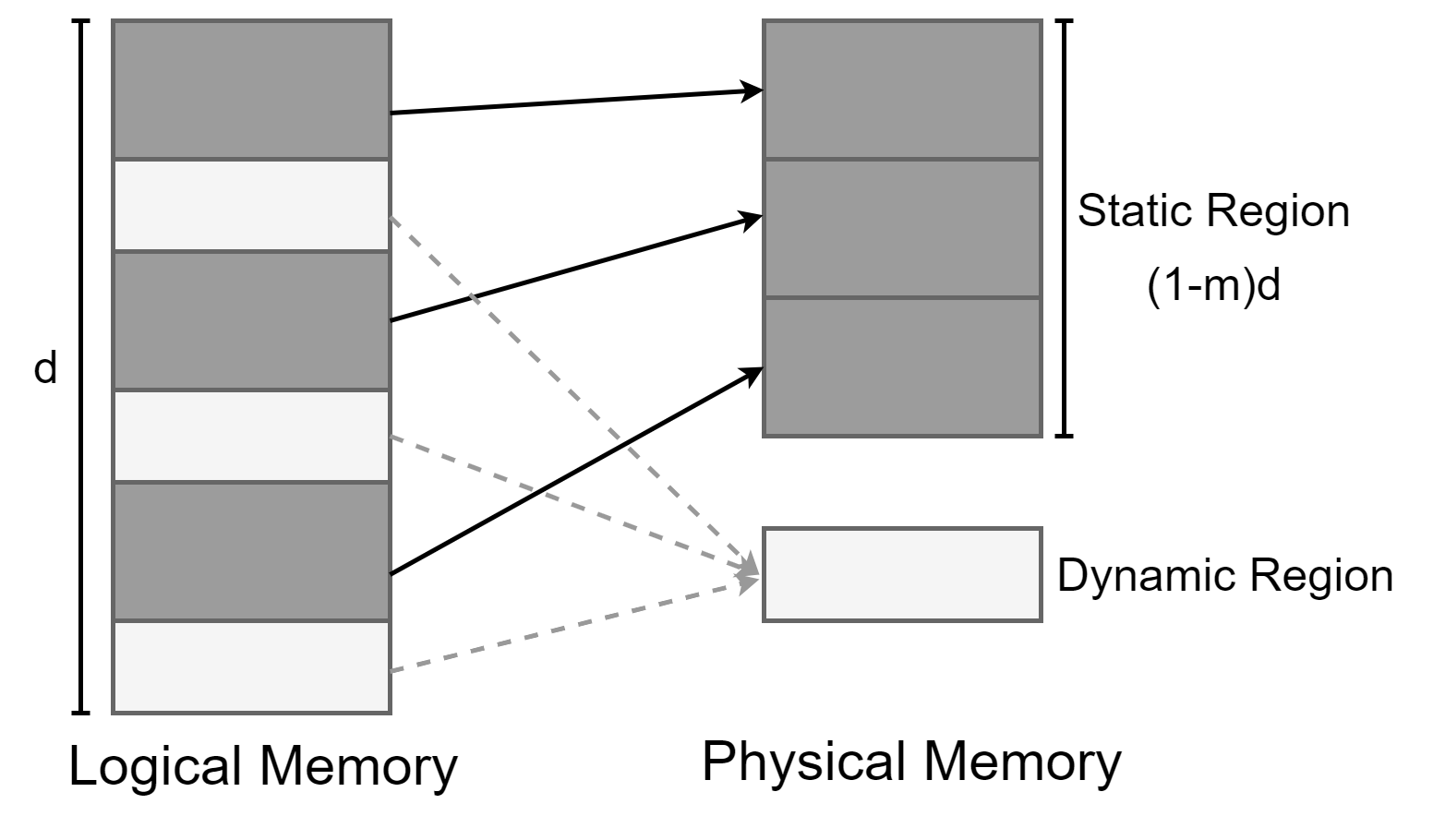}
    \caption{Weight fragmentation. The original weights with depth of $d$ is fragmented into static and dynamic regions. Weights in the dynamic regions are loaded from the off-chip memory at runtime, sharing the same piece of physical memory space in a time-multiplexed manner. The ratio of dynamic regions is denoted as $m$.}
    \label{fig:weight_fragmentation}
\end{figure}

As a common practice in streaming architectures, the convolution and other matrix multiplication operations require the corresponding weight parameters completely stored on-chip. Each operation is assigned with its own dedicated memory block so that these operations can be parallelized and pipelined. As a way to reduce the on-chip memory requirements, a memory fragmentation scheme is introduced, where the original weights memory of depth $d$, is now fragmented into static and dynamic regions, as Figure~\ref{fig:weight_fragmentation} shows. 

Weights under the static regions are fixed and read-only, and these static regions are concatenated together to be stored in a continuous, physical on-chip memory space; while weights in the dynamic regions need to be loaded from the off-chip memory at runtime, sharing the same piece of physical memory space in a time-multiplexed manner. The computational pipeline iteratively consumes the weights from either the static or dynamic regions, controlled by a counter to track the number of reads. 

For each convolution operation, the ratio of its dynamic regions $m \in [0,1]$ is a tunable design parameter, where $m=0$ refers to the situation that all weights are static and no off-chip loading is required, while $m=1$ represents moving everything to off-chip instead. Given this ratio, we can represent the saved on-chip memory depth as

\begin{equation}
    \Delta d = md,
\end{equation}
where $d$ denotes the original memory depth before fragmentation. Meanwhile, the required off-chip memory bandwidth to load the dynamic regions can be represented as the following equation:
\begin{equation}
    \Delta BW = mrc,
\end{equation}
where the computational pipeline consumes the weights at the rate of $r$ (words/cycle), and similar to activations, weights can also be encoded with the compression ratio of $c$ reducing the required bandwidth, but this ratio is compile-time known for weights.

\subsection{Subgraph Reconfiguration}
FPGAs can be reconfigured at runtime, which provides a unique design option for the streaming architecture to partition the entire CNN workload into multiple subgraphs \cite{Venieris2019}. Each subgraph is mapped to a customised hardware design, and these subgraphs are scheduled sequentially on a single device through the device reconfiguration. When using this reconfiguration feature, a trade-off between the latency and the throughput of the system is introduced.

\begin{figure}
    \begin{subfigure}{0.6\columnwidth}
        \centering
        \includegraphics[width=\linewidth]{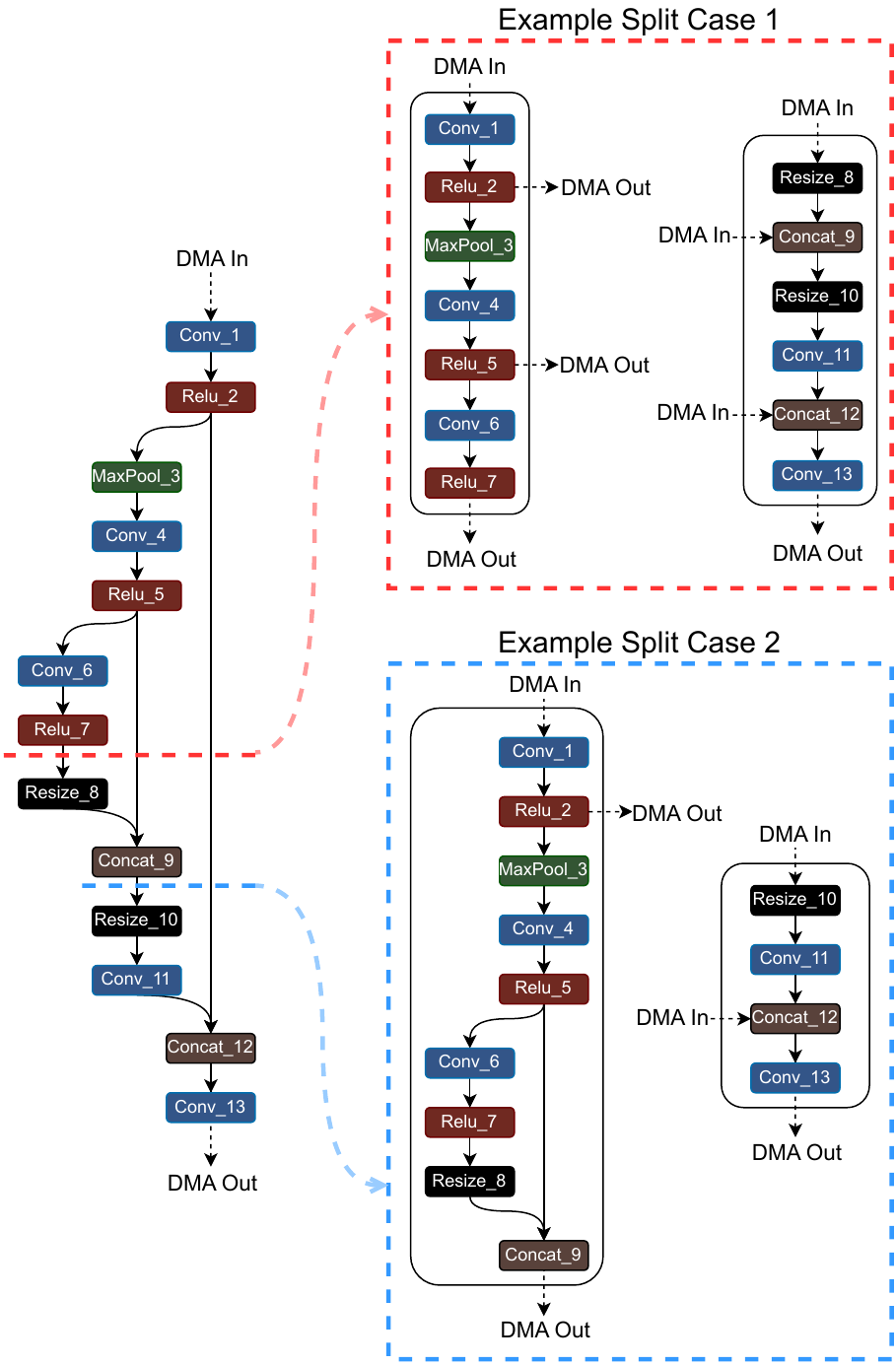}
        \caption{Graph spliting strategies.}
        \label{fig:partitioning_spliting}
    \end{subfigure}
    \hfill %
    \begin{subfigure}{0.3\columnwidth}
        \centering
        \includegraphics[width=\linewidth]{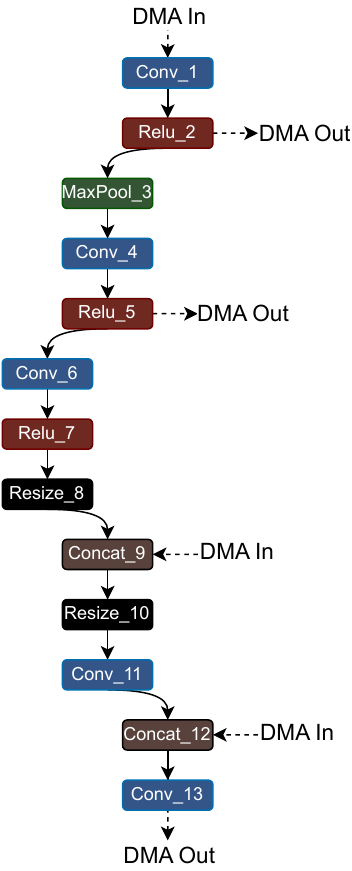}
        \caption{Graph off-chip streaming strategy.}
        \label{fig:partitioning_off_chip_streaming}
    \end{subfigure}
    \caption{Graph manipulation strategies. Figure \ref{fig:partitioning_spliting} showcases the graph partitioning strategy into multiple subgraphs for different reconfiguration points. In Figure \ref{fig:partitioning_off_chip_streaming} the off-chip streaming strategy is presented, utilising the activation eviction method. By streaming the long skip connections to off-chip and reading back from it on merge points the graph can fit into the device without needing multiple partitions and hence device reconfiguration.} 
    \label{fig:partitioning_strategies}
\end{figure}

Assume a targeted CNN is partitioned to $N$ subgraphs, where each subgraph is computed in a batch mode, as multiple image frames are fed into the hardware in a pipelined manner, and the batch size is denoted as $b$. The initial interval, the number of cycles elapsed between the issuing of two consecutive image frames, as well as the depth of the pipeline are denoted as $II_i$ and $d_{pi}$ respectively. After computing the $b$ image frames, the FPGA device is reconfigured with the customised hardware for the next subgraph and the latency penalty for programming the bitstream is denoted as $t_{ri}$. Therefore, the total latency of computing a whole batch of data is represented as follows, assuming the system frequency is $f$.

\begin{equation}
    t = \sum_{i=1}^{N} (b \cdot II_i+d_{pi})/f + N \cdot t_{ri}\;\;\;\;\;\;\;\;(seconds)
    \label{equ:latency}
\end{equation}
The corresponding system throughput is represented as:
\begin{equation}
    \Theta = b/t\;\;\;\;\;\;\;\;\;\;\;\;\;\;\;\;\;\;\;\;\;\;\;\;(frames\;\;per\;\;second)
    \label{equ:throughput}
\end{equation}

In general, a large $N$ means more latency overhead on reconfiguration, but it also may help decrease the $II_i$, since the subgraph becomes smaller and more parallelism can be assigned to each subgraph. Therefore, the choice of $N$ becomes a design variable that needs to be considered.

Figure \ref{fig:partitioning_strategies} illustrates the methods employed to alter the computation graph. We allow graph partitioning or fragmentation/eviction at any position as long as the three constraints listed below are met, 
\begin{itemize}
    \item on-chip resource: Each subgraph's utilization of on-chip FPGA resources, including Digital Signal Processing Blocks (DSP), Look-Up Tables (LUT), Flip-Flops (FF), and Block RAM (BRAM), must adhere to the limits imposed by the FPGA device. 
    \item off-chip bandwidth: The off-chip bandwidth utilization of each subgraph, comprising the subgraph's inputs and outputs, as well as the overhead related to activation eviction and weight fragmentation, must comply with the bandwidth limitations of the FPGA device to prevent bottlenecks or performance degradation.
    \item compute dependency: Subgraphs are scheduled to execute sequentially on a single FPGA device in a time-multiplexed manner. Therefore, it is crucial to maintain the correct sequence of computations. For any vertex within a subgraph, all of its producers must either reside within the same subgraph or have already been executed in a previously scheduled subgraph.
\end{itemize}

Graph partitioning or fragmentation/eviction are not exclusive, and they can be combined together for the optimal performance and resource trade-off, leaving a huge design space to explore. We will elaborate our exploration algorithm in Section~\ref{greedy exploration}.

%% file: Implementation/implementation.tex
\section{Implementation} \label{implementation}

In this section, we present the implementation details of the proposed system architecture, along with the specifics of our Design Space Exploration (DSE) optimization strategy. Additionally, we introduce an updated approach for calculating and modeling the pipeline depth of the computation graph.

\begin{figure*}[ht]
    \centering
    \begin{minipage}{0.75\textwidth}
        \centering
        \includegraphics[width=0.9\linewidth]{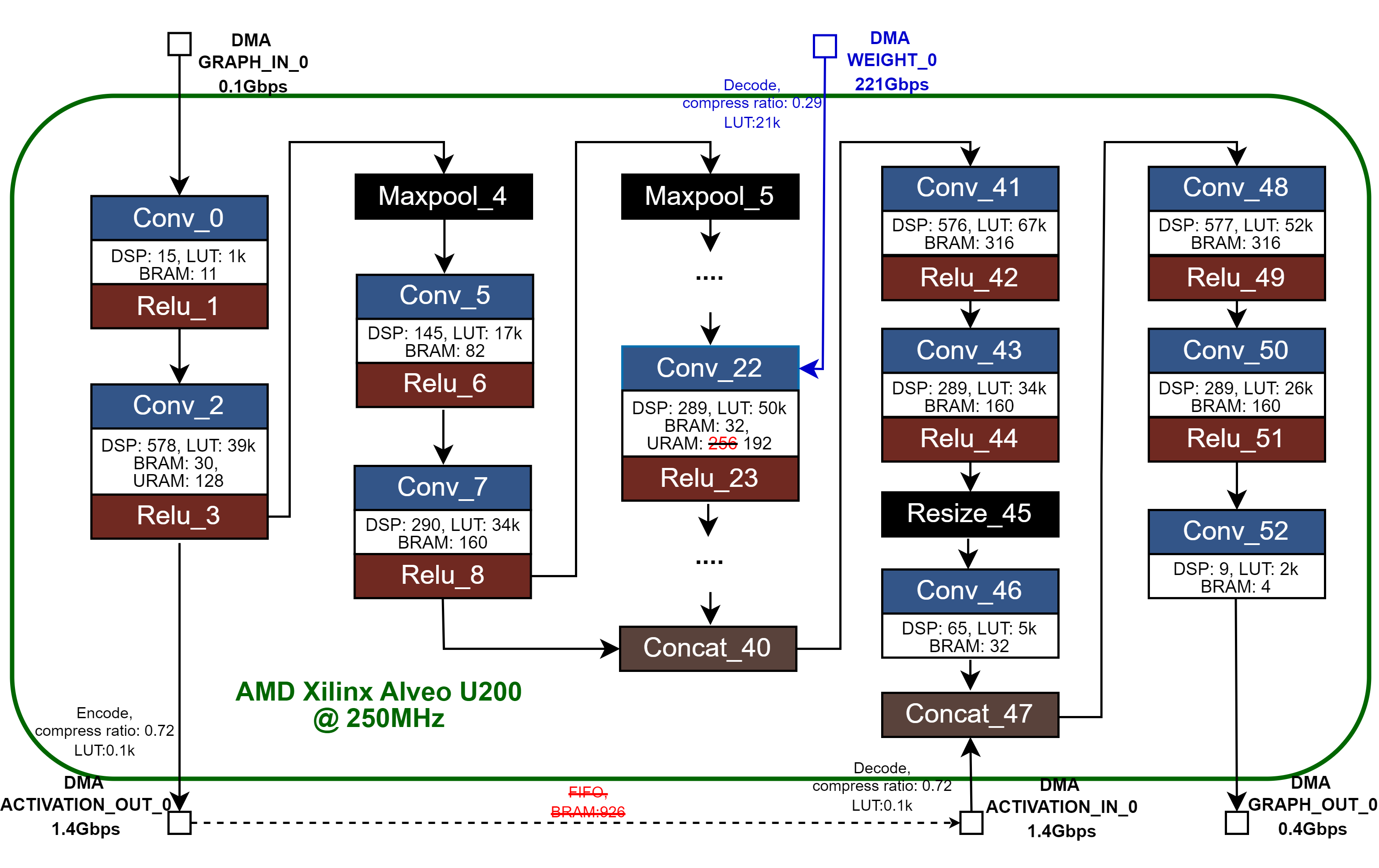}
    \end{minipage}
    \begin{minipage}{0.2\textwidth} %
        \centering
        \footnotesize
        \begin{tabular}{ll}
        \toprule
        model & UNet \\
        batch size & 1 \\
        \midrule
        latency & 47 ms\\
        throughput & 21 fps\\
        \midrule
        DSP & 6062 (89\%)\\
        BRAM & 3654 (85\%) \\
        URAM & 864 (90\%)\\
        LUT & 1040K (88\%)\\
        FF & 841K (36\%)\\
        \midrule
        BW & 225Gbps (37\%)\\
        \bottomrule
        \end{tabular}
    \end{minipage}
    \caption{A streaming accelerator design example with the proposed memory
optimisation methodology, where the weights of the layer {\tt Conv\_22} are partially offloaded to off-chip, and the long skip connection between {\tt Relu\_3} and {\tt Concat\_47} is also evicted to off-chip.}
    \label{fig:example}
\end{figure*}

\subsection{System Architecture} \label{system_architecture}
We incorporate the proposed approach with an open-source toolflow, fpgaConvNet \cite{Venieris2019}. We use fpgaConvNet as a code generator that translates the PyTorch description of the CNN to the RTL files, and stitches those files together to build the computational pipeline. 

Same as other existing streaming architectures, the computation pipeline generated by the original fpgaConvNet toolflow restricts most data access to on-chip memories. When incorporating the activation eviction, we keep the hardware of the computation nodes unchanged and modify their connections as in Figure~\ref{fig:activation_eviction} which exposes two additional DMA ports. In the case of the weight fragmentation, the original on-chip memory block is reduced as in Figure~\ref{fig:weight_fragmentation} and the shared buffer for the dynamic regions is connected to one read-only DMA port. 

The encode/decode modules are inserted to the DMA ports if requested. We currently support both Run-Length Encoding (RLE) and Huffman coding, and they are applied to each data word independently. The resource overhead and the performance impact of these two encoding schemes can be found in Section~\ref{Off-Chip Streaming Compression}. If the total required DMA ports are more than the available DDR banks, time-multiplexing is applied. To avoid routing congestion and degradation of the achievable operating frequency, we also expose
the maximum number of DMA ports as a user-defined parameter whose value can be selected empirically based on the post-placement attempts. 

Figure~\ref{fig:example} presents a UNet design example with the layer-wise computation pipeline and the off-chip memory connections visualized. The input images are fed into the pipeline through {\tt Conv\_0}, and the segmentation masks are generated from {\tt Conv\_52} which is the last layer. As the figure shows, the weight fragmentation is applied to {\tt Conv\_22}, which reduces the URAM usage of this convolutional layer from 256 to 192, saving 18Mb on-chip memory at the cost of 21k LUTs for decoding and 221Gbps bandwidth. In addition, the activation eviction is applied between {\tt Relu\_3} and {\tt Concat\_47} which further saves 926 BRAM, equivalent to 16Mb on-chip memory, at the cost of 2.8Gbps and negligible LUT overhead.

\subsection{DSE and Optimisation} \label{greedy exploration}
To obtain a design like Figure~\ref{fig:example}, there are the following design choices that need to be decided:

\begin{itemize}
    \item \textit{how many subgraphs should we have after partitioning?}
    \item \textit{how to allocate the on-chip resources along the computation pipeline?}
    \item \textit{where the activation eviction should be applied?}
    \item \textit{where the weight fragmentation is used and how to decide the fragmentation ratio?}
\end{itemize}

The above four problems can be formulated into an optimisation problem, where the objective is to maximize the system throughput $\Theta$ (Equation~\ref{equ:throughput}) and minimize the latency $t$ (Equation~\ref{equ:latency}), subjecting to the FPGA device constraints of on-chip resources $R$ and off-chip memory bandwidth $BW$. As such,

\begin{align}
    \max_{\mathbf{D}_v}{\Theta},\min_{\mathbf{D}_v}{t},\;\;\ s.t. \sum_{v} R(v) \leq R, \sum_{v} BW(v) \leq BW \notag \\ \forall v \in S, \forall S \in G 
\end{align}
where $G$, $S$, $v$ represent the graph, subgraph, and the vertex respectively. For each individual vertex, its design choices are denoted by the vector $\mathbf{D}_v$ which contains:
\begin{itemize}
    \item binary flags $s_i, s_o$ indicating if the current vertex is the input or output of a subgraph,
    \item the operation parallelism $p$,
    \item binary flags $a_i, a_o$ indicating whether its inputs or outputs' activations are evicted to off-chip,
    \item weight fragmentation ratio $r$.
\end{itemize}

The process of deciding the values of the vector $\mathbf{D}_v$ is known as Design Space Exploration (DSE). fpgaConvNet proposed a greedy algorithm \cite{montgomerie2022samo} to identify the optimal operation parallelism $p$, and in this work, we extend this algorithm to cover the exploration of the rest design choices that are related to off-chip memory. Specifically, our DSE algorithm includes the following pass:

\textcircled{1} \textit{resource-minimal initialization}: we start the DSE with the configuration that has minimal resource requirement. Specifically, we can partition the target DAG into as many subgraphs as possible, and inside each subgraph, each vertex is also set as it's minimal parallelism level. Users can also constrain the partitioning positions by specifying only the certain types of operation that can become the subgraph inputs/outputs, with the tradeoff between DSE time and quality.

\textcircled{2} \textit{compute parallelism allocation}: for the computation pipeline in a subgraph, it's preferable to increase the parallelism of the slowest operation so that the pipeline initial interval $II_i$ (Equation~\ref{equ:latency}) can be reduced. Whenever the parallelism of the slowest operation cannot be further increased, we will move on to other operations and allocate the compute resource there if it helps reduce the pipeline depth $d_{pi}$.

\textcircled{3} \textit{on-chip memory allocation}: on an AMD Xilinx FPGA device, BRAMs and LUTRAMs are typically available to be the on-chip memory, as well as URAMs on certain devices. These on-chip memory resources are configured to store the weights and activations as required. Our algorithm balances the utilisation ratios among the resource types and also optimises the utilisation efficiency considering that these on-chip memory resources can only be configured to be specific width and depth ratios.

\textcircled{4} \textit{off-chip bandwidth allocation}: this pass decides the binary flags $a_i, a_o$ for activation eviction and also the weight fragmentation ratio $r$. For all the vertices in the subgraph, we sort by the heuristic metric $ L \Delta d/ \Delta BW$, in which $L$ denotes the data word length, while $\Delta d$ and $\Delta BW$ denote the change in the on-chip memory depth and off-chip memory bandwidth respectively. A large quotient indicates a great saving for the on-chip memory with a relatively small cost in off-chip bandwidth, so the bandwidth will be allocated for that vertex in priority. 

\textcircled{5} \textit{partition merging}: since we already start with splitting into DAG into as many subgraphs as possible, this pass focuses on the merging of the subgraphs. When we merge two subgraphs, the total device reconfiguration time is reduced since there are fewer subgraphs, however, the merged subgraph usually has less parallelism due to the device resource constraint, or in an extreme case, the merged subgraph might not fit on the target FPGA device at all. Therefore, this pass examines the performance estimation before and after any potential merging and then takes the decision.

The above passes are combined in an iterative manner, as Algorithm~\ref{algo:dse} shows. Eventually, the DSE process identifies the optimal design configuration which is used to generate the RTL code.

\begin{algorithm}
    \caption{Design Space Exploration}
    \label{algo:dse}
    \begin{algorithmic}
    \Require $G, b$ \Comment{CNN graph, batch size}
    \Require $R, BW, t_{ri}$ \Comment{on-chip rsc, bandwidth, reconf time}
    \State $S \gets$ initialize($G$) \textcircled{1} \Comment{resource minimal subgraphs}
    \For{$S \in G$}
        \While {$\sum_{v} R(v) \leq R$ and $\sum_{v} BW(v) \leq BW$}
            \State alloc\_parallel($S$, $R$) \textcircled{2}
            \State alloc\_on\_chip($S$, $R$) \textcircled{3}
            \State alloc\_off\_chip($S$, $BW$) \textcircled{4}
        \EndWhile
    \EndFor
    \State merge\_subgraph($S_i, S_j$) $\forall S_i, S_j \in G$ \textcircled{5}
    \end{algorithmic}
\end{algorithm}

\subsection{Pipeline Depth Estimation}

Given the importance of pipeline depth on the final performance of a design for a computation graph G, as indicated in Equation \ref{eqn:final_performance}, as well as its impact on memory resources, we developed a revised approach for estimating pipeline depth from the one originally presented in fpgaConvNet.

\begin{table}[h]
    \label{tab:symbols}
    \centering
    \caption{HW related symbols of the model graph $G$}
    \resizebox{\columnwidth}{!}{%
    \begin{tabular}{ll}
        \toprule
        Symbols & Definitions \\
        \midrule
        $r^{in}_v$ & Standard input rate in of vertex $v$\\
        $\sigma^{in}_v$ & Size of the input feature map of vertex $v$\\
        $\rho_v$ & Pipeline depth of vertex $v$\\
        $\lambda_v$ & Latency estimation for vertex $v$\\
        $N_G^{in}$ & The first node of the graph $G$ \\
        \bottomrule
    \end{tabular}
    }
\end{table}

Equation \ref{eqn:prev_interval} presented below outlines the computation of the interval leading up to the point in the graph $G$ where the current vertex $v$ resides. This calculation is mandatory for determining the initiation rate $r^{st}(v)$ of the vertex $v$.
\begin{align} \label{eqn:prev_interval}
    Interval_{prev}(v) = \max(\lambda_a + \rho_a) \forall a \in ancestors(v)
\end{align}
where the $ancestors(v)$ returns all nodes in graph $G$ that have direct connection to the node $v$, and the latency estimation $\lambda_v$ of each graph vertex is derived from the fpgaConvNet's performance models for each respective layer type.

\begin{figure*}[h]
    \centering
    \includegraphics[width=\textwidth, keepaspectratio]{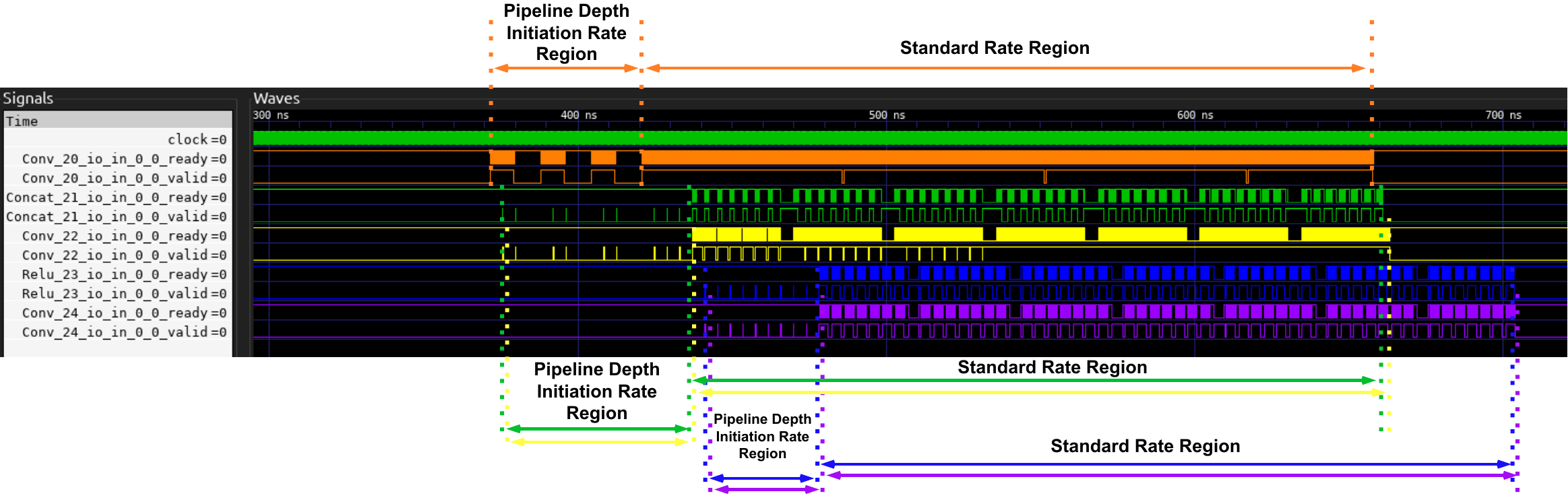}
    \caption{A segment of the design's waveform illustrates the distinction between the initiation rate $r^{st}$, which is applicable to the pipeline depth region of the layers, and the standard input rate $r^{in}$, evident for the remainder of the layer's execution.}
    \label{fig:waveform}
\end{figure*}

The initiation rate calculation is detailed in equation \ref{eqn:initiation_rate}.
\begin{align}  \label{eqn:initiation_rate}
    r^{st}(v)=
    \begin{cases}
        r^{in}_v & ancestors(v) = \emptyset \\
        \frac{\sigma^{in}_v}{Interval_{prev}(v)} & otherwise
    \end{cases}
\end{align}
To facilitate a deeper understanding of this equation and enlighten the practical significance of it, Figure \ref{fig:waveform} is provided. The presented waveform is an annotated segment of a synthesised design, which depicts different computation regions in which the rate of a given layer varies based on the region it resides. Notably, the initiation rate is apparent in the pipeline depth region, indicating a significantly distinct rate than the standard rate that applies to the remainder of the layer's execution.

Finally, the refined pipeline depth of each vertex $v$ is calculated as follows:
\begin{align} \label{eqn:delay_calculation}
    Delay(G, v) = \sum_{n=0}^{max(P_G(N_G^{in}, v))} \frac{\rho_n}{r^{st}(n)}
\end{align}
where the function $P_G(src, trg)$ computes all the valid paths in the graph $G$ originating from the source node ($src$) and terminating at the target node ($trg$).

In conclusion, the equation below presents a refined model for the fpgaConvNet's pipeline depth estimation, as this was previously described in Equation~\ref{equ:latency} with the term $d_{pi}$. The updated estimation accounts for the distinct rates present in each layer's computation. This revised version aligns more closely with the actual hardware implementation, offering a more accurate representation and estimation of the model's performance estimates.
\begin{align} \label{eqn:final_performance}
    d_{pG} = \max_{v \in G}(Delay(G,v))
\end{align}

With the incorporation of this updated pipeline depth estimation model, and consequently, the revised latency model (Equation~\ref{equ:latency}), we have ensured an acceptable deviation level of approximately 12\% on our proposed designs. The extent of this deviation may vary slightly depending on the structure of each model.

%% file: Evaluation/evaluation.tex
\section{Evaluation} \label{evaluation}

This section presents the experimental results of our proposed method's auto-generated accelerators validated in a range of CNN models (both 2D and 3D) targeting a range of AMD FPGA devices such as ZCU102, U200, VCU1525, and VCU118. All the models included in the evaluation are listed in TABLE~\ref{tab:cnn_models}. The selected models have demonstrated exceptional performance across diverse computer vision tasks such as image segmentation (UNet \cite{ronneberger2015u}), object detection (YOLOv8 \cite{redmon2016you, yolov8_ultralytics}), 3D Volumetric Segmentation (UNet3D \cite{cicek2016unet3d}), as well as action recognition (X3D \cite{Feichtenhofer2020}). Aside from supporting a range of computer vision applications and incorporating both 2D and 3D CNNs, the selected models display a wide range of workload and network characteristics, with the unifying feature of integrating long skip connections into their architectures.

\begin{table}[h]
    \centering
    \caption{Characteristics of the evaluated CNN models.}
    \label{tab:cnn_models}
    \resizebox{\columnwidth}{!}{%
    \begin{threeparttable}
    \begin{tabular}{llllllll}
        \hline
        Model & Dataset &
          \begin{tabular}[c]{@{}l@{}}MACs \\ (G)\end{tabular} &
          \begin{tabular}[c]{@{}l@{}}Params\\ (M)\end{tabular} &
          \# Layers &
          \begin{tabular}[c]{@{}l@{}}\# Conv\\ Layers\end{tabular} &
          \begin{tabular}[c]{@{}l@{}}Input Dims\end{tabular} &
          \begin{tabular}[c]{@{}l@{}}Accuracy\\ (\%)\end{tabular} \\ \hline
        Yolov8n & COCO\tnote{$\pm$} \cite{lin2014microsoft} & 4.37 & 3.16 & 115 & 63 & (3, 640, 640) & \begin{tabular}[c]{@{}l@{}}37.10 (f32)\\ 37.10 (fp16)\\ 29.60 (fp8)\\ 35.10 (bfp8)\end{tabular} \\ \hline
        UNet    & CamVid\tnote{$\ddagger$} \cite{brostow2009semantic} & 130.12 & 28.96 & 53 & 23 & (3, 368, 480) & \begin{tabular}[c]{@{}l@{}}71.67 (f32)\\ 71.67 (fp16)\\ 60.62 (fp8)\\ 71.75 (bfp8)\end{tabular} \\ \hline
        UNet3D  & BraTS2020\tnote{$\mp$} \cite{menze2014multimodal} & 918.64 & 5.65 & 52 & 19 & (4, 155, 240, 240) & \begin{tabular}[c]{@{}l@{}}85.34 (f32)\\ 85.23 (fp16)\\ 1.15 (fp8)\\ 85.34 (bfp8)\end{tabular} \\ \hline
        X3D-M   & UCF101\tnote{$\dagger$} \cite{soomro2012ucf101} & 6.97 & 3.82 & 396 & 115 & (3, 16, 256, 256) & \begin{tabular}[c]{@{}l@{}}96.40 (f32)\\ 96.40 (fp16)\\ 0.81 (fp8)\\ 96.29 (bfp8)\end{tabular} \\ \hline
    \end{tabular}%
    \begin{tablenotes}
        \item[$\pm$] Accuracy reported in mAP50-95.
        \item[$\ddagger$] Accuracy reported in mIOU.
        \item[$\mp$] Accuracy reported in Dice Score.
        \item[$\dagger$] Accuracy reported in Top-1 Acc.
    \end{tablenotes}
    \end{threeparttable}
    }
\end{table}

\subsection{Baseline Setup} \label{baseline_setup}

In this section, we briefly describe the baseline configuration for the ablation study of the next sections. To meet the storage requirements, we utilise quantisation to reduce on-chip resource demands. Weights are quantized depending on minimum and maximum values within each channel of each layer, using a channel-wise block floating-point format with an 8-bit wordlength. Without additional fine-tuning, this process is applied directly to pre-trained weights. Activations are likewise quantized with an 8-bit fixed wordlength, based on a calibration set of data to capture their range. The quantisation approach has a modest accuracy drop, as shown in TABLE~\ref{tab:cnn_models}, while providing significant improvements in resource utilisation and model performance. Moreover, we leverage the device's URAM, whenever it is supported, to balance the utilisation ratios between BRAM and URAM, thereby optimising overall utilisation efficiency as detailed in Section \ref{system_architecture}.

The reconfiguration of the device and partitioning of the models provide further enhancements to the baseline configuration. As shown in TABLE~\ref{tab:reconf}, properly partitioning the models across the available resources on the FPGA allows for larger models to be deployed efficiently. This partitioning strategy, supplementary to the quantisation and URAM utilisation techniques, enables the mapping of models that would otherwise exceed the on-chip memory capacity. From this point onward, the term "baseline" will refer to the designs incorporating all the features mentioned in this section, including quantisation, URAM utilisation, and model partitioning.

\begin{table}[h]
    \centering
    \caption{Breakdown and impact of model partitioning with device reconfiguration on UNet3D designs targeting different batch sizes.}
    \label{tab:reconf}
    \resizebox{\columnwidth}{!}{%
    \begin{threeparttable}
    \begin{tabular}{lcccc}
        \hline
        Batch Size                            & 1     & 4     & 16   & 64    \\
        \hline
        Number of partitions                  & 4     & 5     & 6    & 6     \\
        Batch Latency (sec)                   & 0.77  & 2.76  & 9.54 & 36.69 \\
        Batch Compute Time (sec)              & 0.53  & 2.43  & 9.13 & 36.28 \\
        Batch Reconfig Time (sec)             & 0.24  & 0.33  & 0.41 & 0.41  \\
        \hline
        \begin{tabular}[c]{@{}l@{}}Reconfig Contribution on\\ Batch Latency (\%)\end{tabular} & 31.16 & 11.95 & 4.29 & 1.11  \\ \hline
    \end{tabular}%
    \end{threeparttable}
    }
\end{table}

The primary focus of the ablation study is on two models: Unet, a 2D model, and Unet3D, a 3D model. However, it is essential to note that the conclusions derived from this comprehensive analysis are applicable and hold true to the other models evaluated in the final section.
 
\subsection{Off-Chip Streaming}

\begin{figure}[t]
    \centering
    \begin{subfigure}[t]{0.85\columnwidth}
        \includegraphics[width=\linewidth]{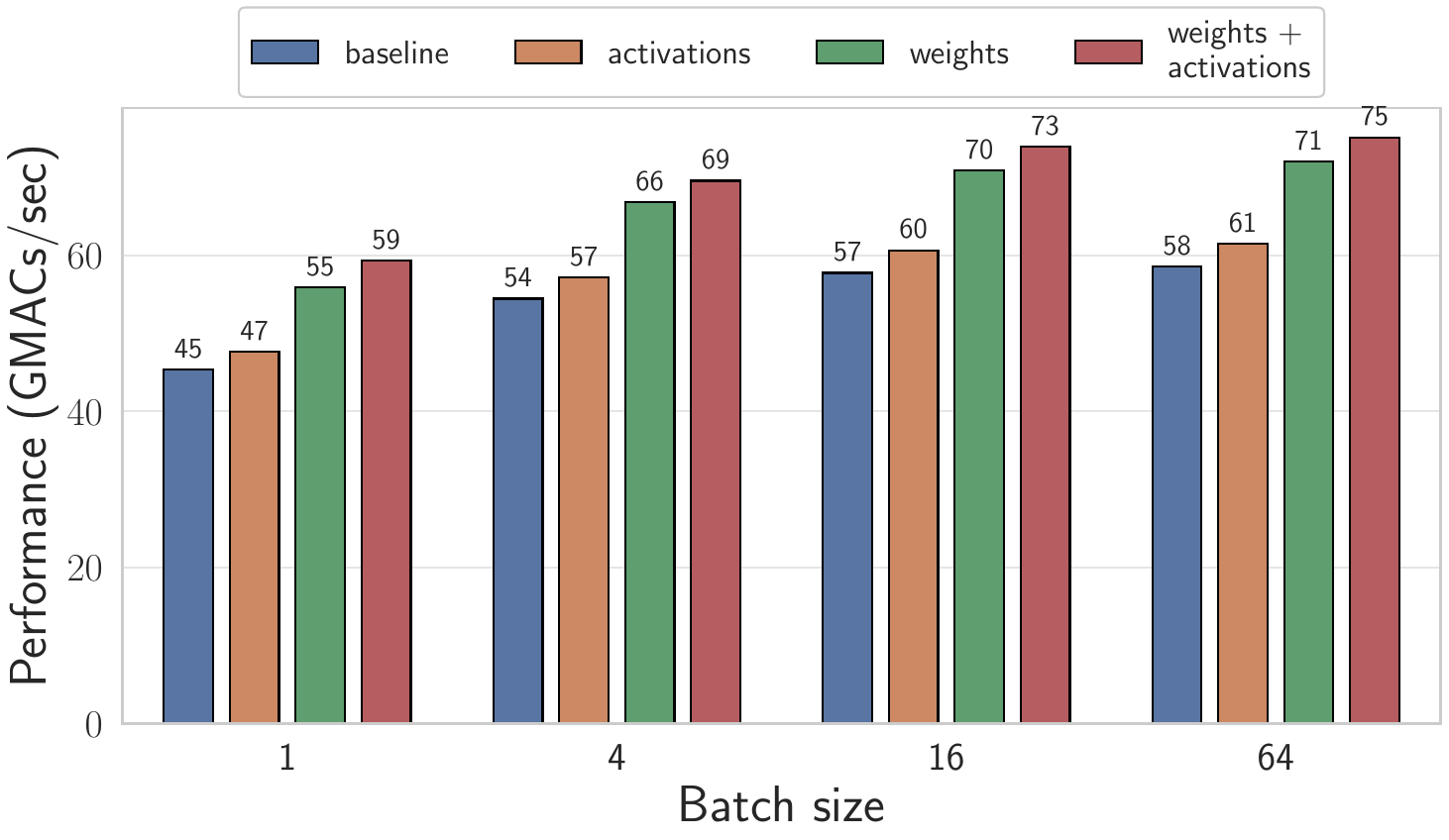}
        \caption{UNet performance under different off-chip streaming strategies}
        \label{fig:off_chip_streaming_unet}
    \end{subfigure}%
    \vspace{\baselineskip} %
    \begin{subfigure}[b]{0.85\columnwidth}
        \includegraphics[width=\linewidth]{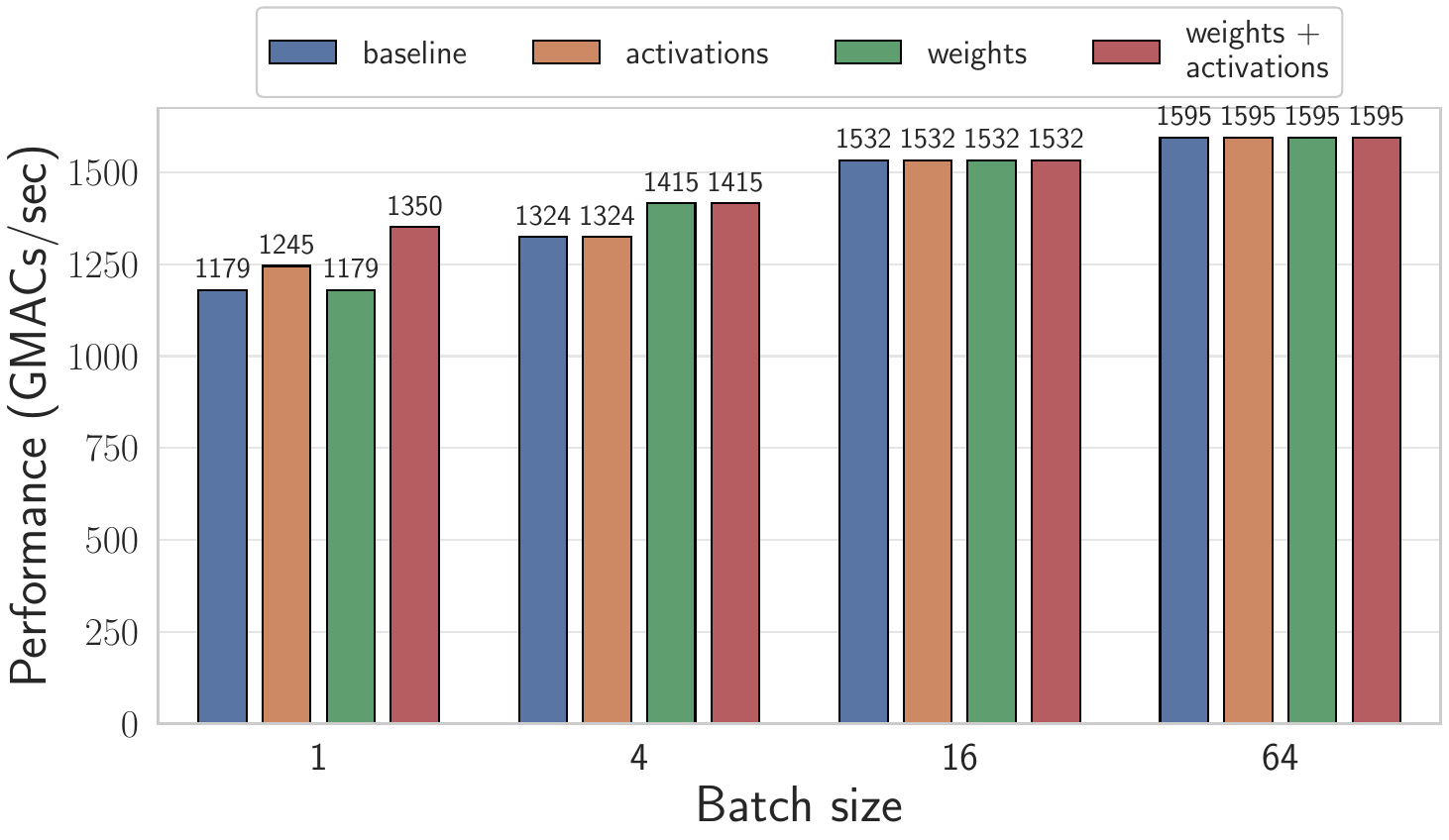}
        \caption{UNet3D performance under different off-chip streaming strategies}
        \label{fig:off_chip_streaming_unet3d}
    \end{subfigure}
    \caption{Evaluation of the UNet and UNet3D under four different steaming strategies: $i)$ No off-chip streaming allowed (\textcolor{blue}{blue}), $ii)$ Only activations off-chip streaming allowed (\textcolor{orange}{orange}), $iii)$ Only weights off-chip streaming allowed (\textcolor{green}{green}), $iv)$ Both weights and activations off-chip streaming (\textcolor{red}{red}).}
    \label{fig:off_chip_streaming}
\end{figure}

In Section~\ref{methodology}, we provide both activation eviction and weight fragmentation to overcome the on-chip memory bottleneck. To better understand their individual contribution to the overall performance, we conduct an ablation study in Figure~\ref{fig:off_chip_streaming}, and we have the following observations:

\begin{itemize}
    \item The \textcolor{blue}{blue} bars are the baseline designs, as these described in Section\ref{baseline_setup}, which have no support for activation eviction and weight fragmentation at all.
    \item When enabling activation eviction only (\textcolor{orange}{orange} bars), compared with the baseline (\textcolor{blue}{blue} bars), there is a small performance improvement for the UNet model of $\sim5\%$. In the case of UNet3D with a batch size of 1, there is a notable improvement in performance, increasing from 1179 to 1245 MACs/sec. 
    \item Moreover, in the case of UNet, there are performance improvements over both the previous cases when using weight fragmentation only (\textcolor{green}{green} bars). The same does not apply for the UNet3D case where the weight fragmentation only performs the same as the baseline approach.
    \item For both the cases of UNet and UNet3D (for batch size 1) the use of both methodologies (\textcolor{red}{red} bars) demonstrate significant improvement in performance up to 1.3$\times$ for UNet and 1.1$\times$ for UNet3D. These results showcase the potential performance gains when the activation eviction and weight fragmentation are enabled under different model workloads and FPGA devices.
    \item We also found the benefits of activation eviction and weight fragmentation are more obvious under smaller batch sizes. For a relatively large batch size such as 64, our DSE algorithm tends to create more subgraphs instead of using activation eviction and weight fragmentation methodologies. This is due to the overhead of subgraph reconfiguration decreases relatively, as Equation~\eqref{equ:throughput} shows.
\end{itemize}

\subsection{Off-Chip Streaming Compression}
\label{Off-Chip Streaming Compression}

\begin{figure}[t]
    \centering
    \begin{subfigure}[t]{0.85\columnwidth}
        \includegraphics[width=\linewidth]{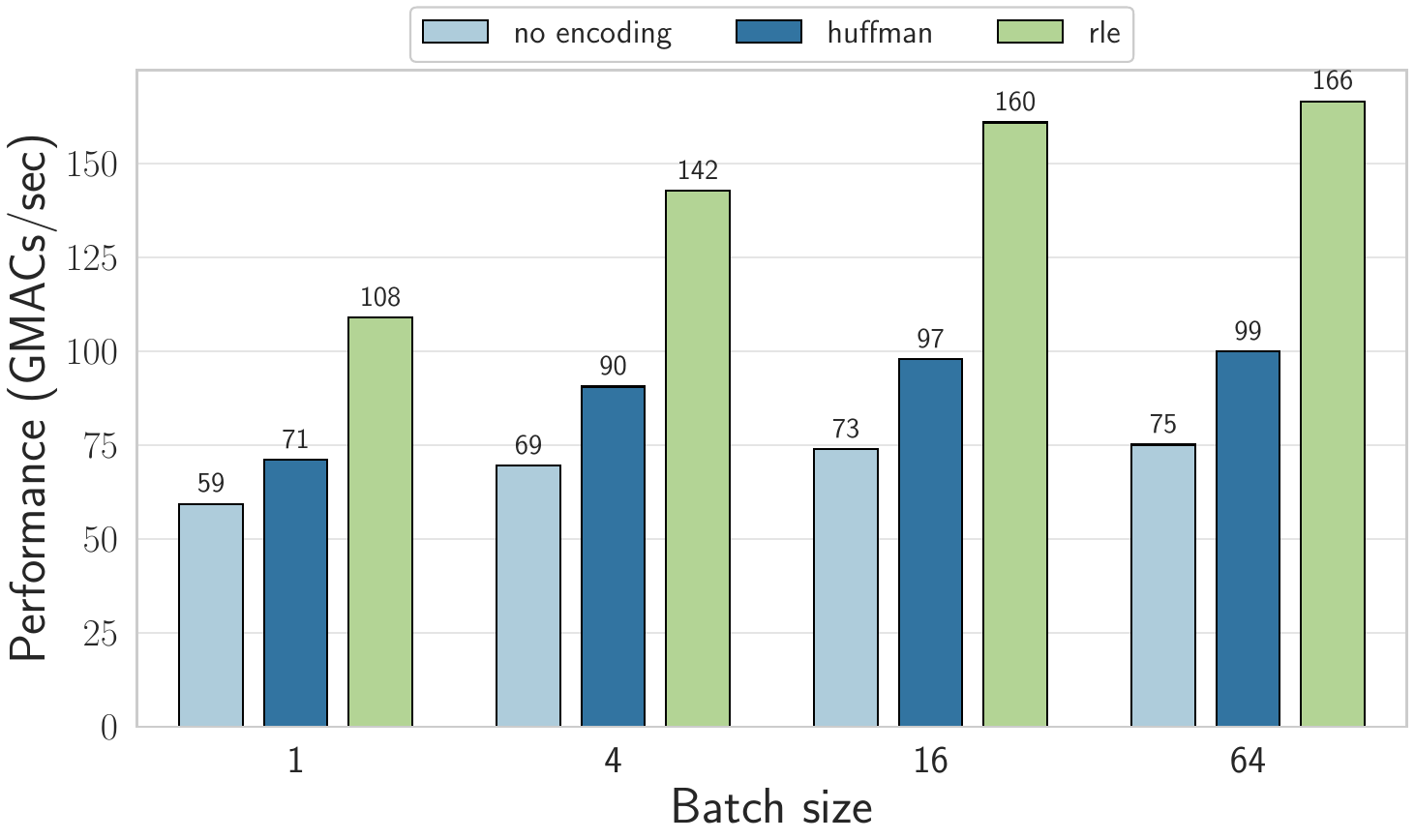}
        \caption{UNet performance under different compression strategies}
        \label{fig:compression_unet}
    \end{subfigure}%
    \vspace{\baselineskip} %
    \begin{subfigure}[b]{0.85\columnwidth}
        \includegraphics[width=\linewidth]{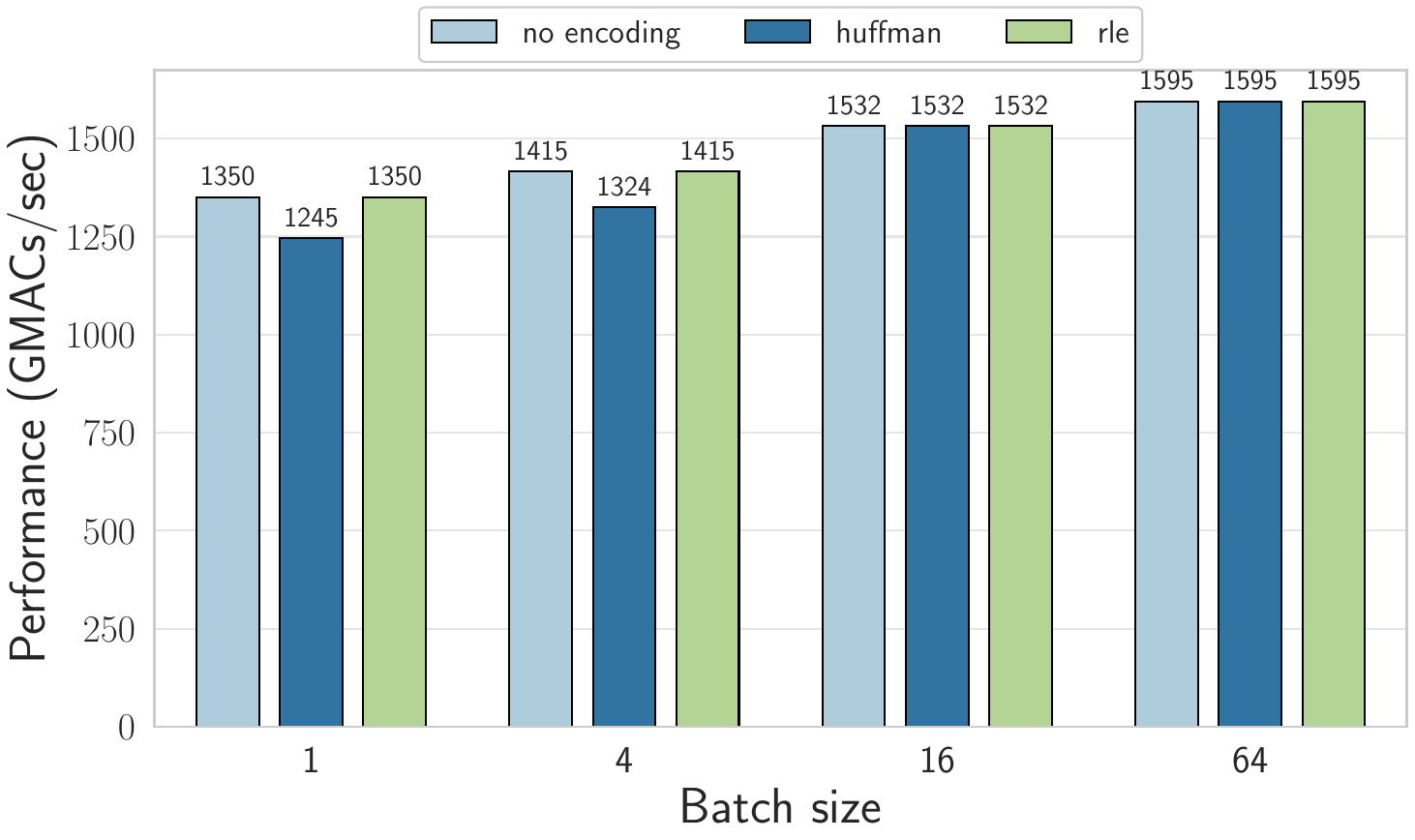}
        \caption{UNet3D performance under different compression strategies}
        \label{fig:compression_unet3d}
    \end{subfigure}
    \caption{Evaluation of the UNet and UNet3D under three compression strategies for off-chip streaming. The off-chip streaming configuration is fixed to \textit{weights \& activations streaming}. The different colours denote the encoding scheme adopted: \textit{No encoding}, \textit{Huffman encoding}, and \textit{run-length encoding (RLE)}.}
    \label{fig:compression}
\end{figure}

To efficiently utilise the off-chip memory bandwidth, we transfer the data in the encoded formats such as Huffman and run-length encoding (RLE). The performance impact of those encoding schemes is shown in Figure~\ref{fig:compression}. The resource overhead, which includes a fixed encoding and decoding cost in LUTs and FFs per data stream, increases proportionally with the number of parallel data streams employed.

\begin{table*}[t]
\centering
\caption{
Comparison of FPGA accelerator architectures generated by SMOF with prior state-of-the-art works on four major computer vision tasks (segmentation, detection, 3D segmentation, action recognition).
}
\label{tab:fpga-comparison}
\resizebox{\textwidth}{!}{%
\begin{threeparttable}
\begin{tabular}{@{}rcccc|ccc||c|c||c|c||c@{}}

\toprule
                            &  S. Basalama \cite{basalama2023flexcnn} & S. Liu \cite{liu2018optimizing} & S. Liu \cite{liu2019towards} & Vitis-AI \cite{kathail2020xilinx} & \multicolumn{3}{c||}{SMOF (Ours)} & A. Montgomerie \cite{montgomerie2023satay} & SMOF (Ours) & P. Toupas \cite{toupas2023fpgahart} & SMOF (Ours) & SMOF (Ours)\\
\midrule
\midrule
Model                       & UNet\tnote{$\dagger$}    & UNet\tnote{$\dagger$}            & UNet\tnote{$\dagger$}            & UNet\tnote{$\dagger$}          & \multicolumn{3}{c||}{UNet}      & YOLOv8s      & YOLOv8n       & X3D-M     & X3D-M    & UNet3D  \\

Model Type                  & 2D                        & 2D              & 2D              & 2D                & \multicolumn{3}{c||}{2D}    & 2D          & 2D           & 3D       & 3D        & 3D      \\ 
Input Size                  & $256\times256$            & $512\times512$  & $256\times256$  & $128\times128$    & \multicolumn{3}{c||}{$256\times256$}      & $640\times640$          & $640\times640$           & $16\times256\times256$      & $16\times256\times256$      & $155\times240\times240$     \\ 
MACs                        & 12                        & 5.9             & 27.4            & 5.6           & \multicolumn{3}{c||}{130}       & 30.5         & 4.4         & 6.97      & 6.97     & 918.6  \\ 
Precision                   & Float32                   & W16A16          & W8A8            & W16A16        & \multicolumn{3}{c||}{W8A8}      & W8A16        & W8A8        & W16A16    & W8A8     & W8A8    \\ 
\midrule
Device                      & VCU1525                   & ZC7045          & Arria 10        & ZCU102        & VCU1525   & U200  & ZCU102    & VCU118       & VCU118        & ZCU102    & ZCU102   & U200    \\
Reconfigurations\tnote{$\mp$}            & -                         & -               & -               & -             & -         & -       & 6       & -            & 5             & -         & 9        & 5    \\
LUT                         & 567K                      & 85K             & 170K (ALM)      & 52K           & 993K      & 1040K    & 213K   & 1023K        & 543K          & -         & 235K     & 289K    \\
DSP                         & 3078                      & 640             & 1665            & 710           & 6019      & 6062    & 1461    & 6815         & 5061          & 2116      & 932      & 5677    \\
BRAM (18K)                  & 1598                      & 545             & 21894 (M20K)    & 510           & 3715      & 3654    & 1368    & 1322         & 1813          & 948       & 857      & 2980    \\
URAM                        & 422                       & -               & -               & -             & 864       & 864     & -      & 713          & 431           & -         & -        & 528     \\
Freq. (MHz)                 & 234                       & 200             & 200             & 281           & 200       & 250     & 200     & 240          & 250           & 160       & 200      & 250     \\
\midrule
Througput (fps)\tnote{$\ddagger$}  & 17.18              & 17.12           & 57.47           & 156.7         & 16.96     & 21.21   & 1.28\tnote{$\pm$}   & 8.33         & 184.27\tnote{$\pm$}        & 13.44\tnote{$\pm$}     & 27.08\tnote{$\pm$}    & 1.75\tnote{$\pm$}    \\  
GOP/s                              & 207                & 107             & 1578            & 832           & 2206      & 2758    & 166    & 1244       & 808          & 86        & 171     & 1595    \\ 
GOP/s/DSP                          & 0.06               & 0.16            & 0.94            & 0.22          & 0.36      & 0.45    & 0.11    & 0.18      & 0.16         & 0.03      & 0.18    & 0.28    \\ 

\bottomrule
\end{tabular}
\begin{tablenotes}
    \item[$\dagger$] Optimised UNet architectures tailored to the HW design (variations in MACs)
    \item[$\ddagger$] For 3D CNNs the throughput is defined as \textit{clips/s} instead
    \item[$\pm$] Results on designs with multiple partitions reported with favourable batch size
    \item[$\mp$] Designs with `-' indicate no extra reconfiguration of the device after the initial bitstream in loaded
\end{tablenotes}
\end{threeparttable}
}
\end{table*}

In order to evaluate the proposed encoding schemes, we apply the off-chip streaming configuration outlined in the preceding section, which enables weight fragmentation and activation eviction. For UNet (Figure~\ref{fig:compression_unet}), we observe the RLE scheme is the best choice, providing up to 2.21$\times$ MACs/sec compared with the non-encoding case, and 1.68$\times$ MACs/sec compared with the Huffman case. However, in the case of UNet3D (Figure~\ref{fig:compression_unet3d}), applying these encoding schemes does not provide any further performance gain, which is because that the design becomes LUT-bounded instead of BRAM/bandwidth-bounded. In fact, when using the Huffman encoding for UNet3D, the performance actually drops slightly due to the resource overhead for encoding and decoding, making the LUT-bounded situation even worse in this case. Overall, we found offloading weights and activations to the off-chip memory is beneficial to performance, but whether to further apply any encoding scheme may vary case by case.

\begin{figure}[h]
    \centering
    \includegraphics[width=0.9\columnwidth, keepaspectratio]{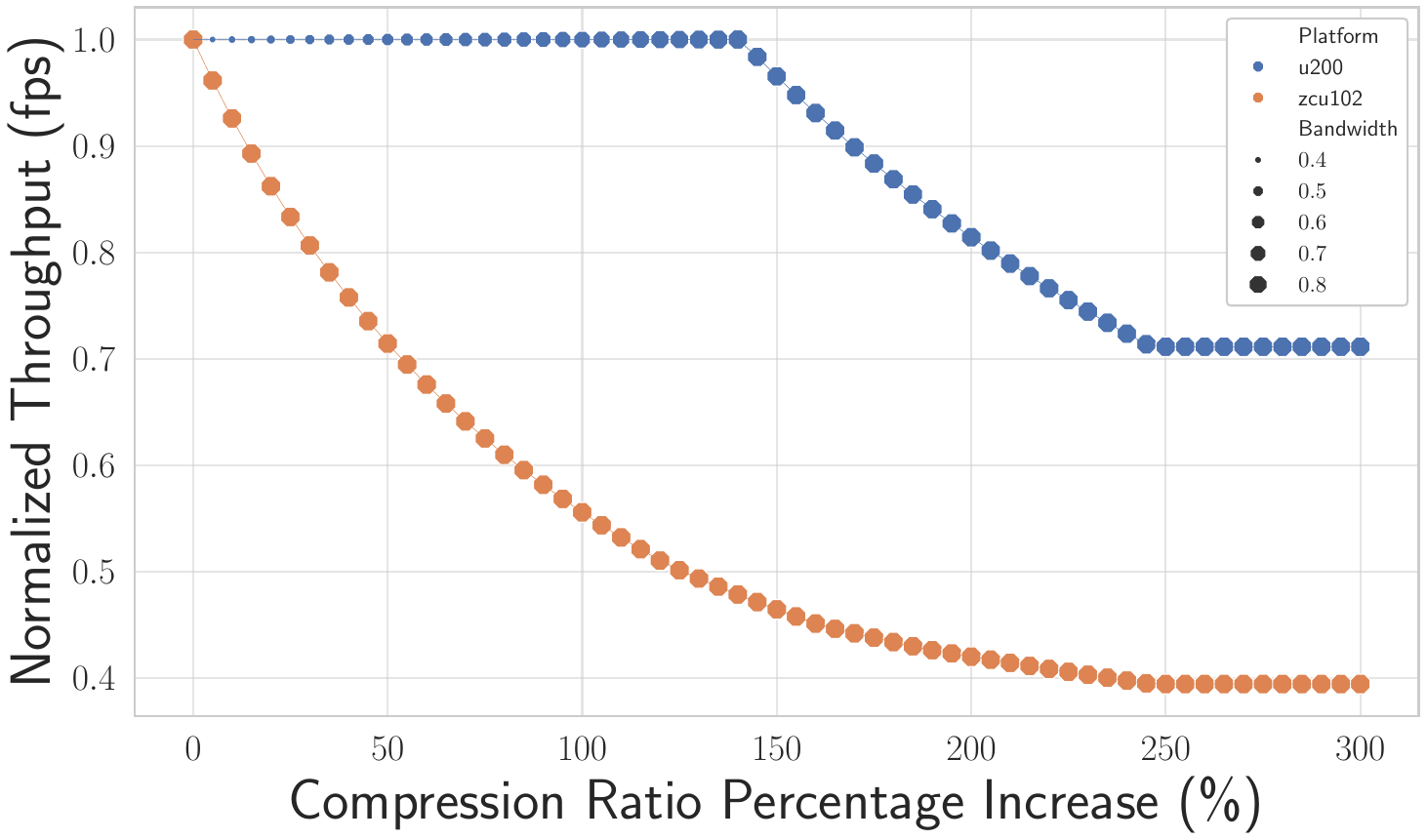}
    \caption{Impact of runtime variability in the compression ratio of activation data on the design's performance due to memory bandwidth limitations.}
    \label{fig:compression_ratio_scaling}
\end{figure}

Another noteworthy aspect is the potential runtime variability in the compression ratio of activation data, impacting performance as illustrated in Figure \ref{fig:compression_ratio_scaling}. In this experiment, we incrementally increase the compression ratio so that more bandwidth is actually required than the predicted one. When this kind of bandwidth under-estimation happens, if there is still some leftover bandwidth on the device to accommodate the difference, there is no performance degradation, so the line of U200 plateaus when the compression ratio difference is less than 140\%. However, when the leftover bandwidth is insufficient to cover this difference, undesired stalls occur during computation, resulting in performance degradation as depicted by the curves.

\subsection{Comparison with related FPGA works}

TABLE~\ref{tab:fpga-comparison} presents a thorough comparison of the accelerator designs produced by our method with those from previous studies. The evaluation includes a range of 2D and 3D Convolutional Neural Network (CNN) models on several FPGA platforms. 
Our solution has a significant advantage due to its adaptability. Our designs exhibit the ability to produce a broad spectrum of accelerators tailored to the characteristics of each one of the four major computer vision applications that we target: action recognition, detection, 3D segmentation, and segmentation. This adaptability extends across both 2D and 3D spatial domains, showcasing the adaptability of our approach in addressing a wide spectrum of applications.

In regards to image segmentation using the UNet, prior studies improved and optimised the model architecture, resulting in a reduction of the overall number of operations of the original model. Variations in input size also affect the total MAC count. For this reason, we provide three metrics: \textit{fps, GOP/s, and GOP/s/DSP}, to account for variations in MACs as well as device diversity.

In comparison to \cite{basalama2023flexcnn}, our UNet solution on VCU1525 demonstrates similar fps performance. However, our design showcases a notable 10.65$\times$ improvement when normalised over the total operations of the model using the GOP/s metric. Due to the ZCU102's limited resources and the large workload of our UNet model (130 MACs versus 5.6 in \cite{kathail2020xilinx}, a 23.3X increase in model workload), our design becomes compute bounded on this device, resulting in no performance improvement over \cite{kathail2020xilinx}. However, targeting a larger device with more resources, such as the U200, our design demonstrates a 2$\times$ improvement in GOP/s/DSP over \cite{kathail2020xilinx}. This metric, as explained earlier, is device-agnostic as it normalises over the DSPs. We likewise achieve a 2.8$\times$ improvement over \cite{liu2018optimizing} under the same metric and device. Despite achieving 1.7$\times$ greater GOP/s performance than \cite{liu2019towards}, our solution achieves half the GOP/s/DSP, which is likely owing to DSP implementation variations between Intel and AMD devices.

In object recognition task with YOLOv8, our proposed method achieves a 23$\times$ fps improvement over \cite{montgomerie2023satay}. However, it exhibits a slightly lower GOP/s/DSP performance of 0.16 compared to the 0.18 reported in \cite{montgomerie2023satay}. Transitioning to 3D CNNs and the task of action recognition, we accomplish a 1.98$\times$ improvement in GOP/s under the same setup as in [30]. No prior work has been identified for the 3D segmentation task using UNet3D. Our proposed design achieved a performance of 1595 GOP/s targeting the U200 FPGA device.

%% file: Conclusion/conclusion.tex
\section{Conclusion} \label{conclusion}
This paper addresses the challenges in scaling CNNs with complex connections and substantial parameters when mapped onto FPGAs. While streaming-based dataflow architectures have demonstrated a great potential, the conventional approach involves restricting the majority of data movements within the FPGA chip. This restriction frequently becomes a limitation in generating architectures of this nature. To resolve the on-chip memory bottleneck, we present a novel memory optimisation methodology that strategically allocates and utilises both on-chip and off-chip memory within a layerwise pipelined, streaming architecture. Our methodology is capable of partially offloading weights and activations to off-chip memory without penalising the computation pipeline. Additionally, our subgraph partitioning methodology provides a flexible trade-off between latency and throughput, leveraging the reconfigurability of FPGAs. We demonstrate competitive and even state-of-the-art performance, especially on networks with complex, hierarchical skip connections.

%% file: bare_conf.bbl
\begin{thebibliography}{10}
\providecommand{\url}[1]{#1}
\csname url@samestyle\endcsname
\providecommand{\newblock}{\relax}
\providecommand{\bibinfo}[2]{#2}
\providecommand{\BIBentrySTDinterwordspacing}{\spaceskip=0pt\relax}
\providecommand{\BIBentryALTinterwordstretchfactor}{4}
\providecommand{\BIBentryALTinterwordspacing}{\spaceskip=\fontdimen2\font plus
\BIBentryALTinterwordstretchfactor\fontdimen3\font minus \fontdimen4\font\relax}
\providecommand{\BIBforeignlanguage}[2]{{%
\expandafter\ifx\csname l@#1\endcsname\relax
\typeout{** WARNING: IEEEtran.bst: No hyphenation pattern has been}%
\typeout{** loaded for the language `#1'. Using the pattern for}%
\typeout{** the default language instead.}%
\else
\language=\csname l@#1\endcsname
\fi
#2}}
\providecommand{\BIBdecl}{\relax}
\BIBdecl

\bibitem{gokhale2017snowflake}
V.~Gokhale, A.~Zaidy, A.~X.~M. Chang, and E.~Culurciello, ``Snowflake: A model agnostic accelerator for deep convolutional neural networks,'' \emph{arXiv preprint arXiv:1708.02579}, 2017.

\bibitem{guo2017angel}
K.~Guo, L.~Sui, J.~Qiu, J.~Yu, J.~Wang, S.~Yao, S.~Han, Y.~Wang, and H.~Yang, ``Angel-eye: A complete design flow for mapping cnn onto embedded fpga,'' \emph{IEEE transactions on computer-aided design of integrated circuits and systems}, vol.~37, no.~1, pp. 35--47, 2017.

\bibitem{Venieris2019}
S.~I. Venieris and C.~S. Bouganis, ``{FpgaConvNet: Mapping Regular and Irregular Convolutional Neural Networks on FPGAs},'' \emph{IEEE Transactions on Neural Networks and Learning Systems}, vol.~30, no.~2, 2019.

\bibitem{umuroglu2017finn}
Y.~Umuroglu, N.~J. Fraser, G.~Gambardella, M.~Blott, P.~Leong, M.~Jahre, and K.~Vissers, ``Finn: A framework for fast, scalable binarized neural network inference,'' in \emph{Proceedings of the 2017 ACM/SIGDA international symposium on field-programmable gate arrays}, 2017, pp. 65--74.

\bibitem{fahim2021hls4ml}
F.~Fahim, B.~Hawks, C.~Herwig, J.~Hirschauer, S.~Jindariani, N.~Tran, L.~P. Carloni, G.~Di~Guglielmo, P.~Harris, J.~Krupa \emph{et~al.}, ``hls4ml: An open-source codesign workflow to empower scientific low-power machine learning devices,'' \emph{arXiv preprint arXiv:2103.05579}, 2021.

\bibitem{hall2020tensorflow}
M.~Hall and V.~Betz, ``From tensorflow graphs to luts and wires: Automated sparse and physically aware cnn hardware generation,'' in \emph{2020 International Conference on Field-Programmable Technology (ICFPT)}.\hskip 1em plus 0.5em minus 0.4em\relax IEEE, 2020, pp. 56--65.

\bibitem{ronneberger2015u}
O.~Ronneberger, P.~Fischer, and T.~Brox, ``U-net: Convolutional networks for biomedical image segmentation,'' in \emph{Medical Image Computing and Computer-Assisted Intervention--MICCAI 2015: 18th International Conference, Munich, Germany, October 5-9, 2015, Proceedings, Part III 18}.\hskip 1em plus 0.5em minus 0.4em\relax Springer, 2015, pp. 234--241.

\bibitem{redmon2016you}
J.~Redmon, S.~Divvala, R.~Girshick, and A.~Farhadi, ``You only look once: Unified, real-time object detection,'' in \emph{Proceedings of the IEEE conference on computer vision and pattern recognition}, 2016, pp. 779--788.

\bibitem{Feichtenhofer2020}
C.~Feichtenhofer, ``{X3D: Expanding Architectures for Efficient Video Recognition},'' in \emph{IEEE/CVF Conference on Computer Vision and Pattern Recognition}, 2020.

\bibitem{fowers2018configurable}
J.~Fowers, K.~Ovtcharov, M.~Papamichael, T.~Massengill, M.~Liu, D.~Lo, S.~Alkalay, M.~Haselman, L.~Adams, M.~Ghandi \emph{et~al.}, ``A configurable cloud-scale dnn processor for real-time ai,'' in \emph{2018 ACM/IEEE 45th Annual International Symposium on Computer Architecture (ISCA)}.\hskip 1em plus 0.5em minus 0.4em\relax IEEE, 2018, pp. 1--14.

\bibitem{shen2018towards}
J.~Shen, Y.~Huang, Z.~Wang, Y.~Qiao, M.~Wen, and C.~Zhang, ``Towards a uniform template-based architecture for accelerating 2d and 3d cnns on fpga,'' in \emph{Proceedings of the 2018 ACM/SIGDA International Symposium on Field-Programmable Gate Arrays}, 2018, pp. 97--106.

\bibitem{liu2018optimizing}
S.~Liu, H.~Fan, X.~Niu, H.-c. Ng, Y.~Chu, and W.~Luk, ``Optimizing cnn-based segmentation with deeply customized convolutional and deconvolutional architectures on fpga,'' \emph{ACM Transactions on Reconfigurable Technology and Systems (TRETS)}, vol.~11, no.~3, pp. 1--22, 2018.

\bibitem{kathail2020xilinx}
V.~Kathail, ``Xilinx vitis unified software platform,'' in \emph{Proceedings of the 2020 ACM/SIGDA International Symposium on Field-Programmable Gate Arrays}, 2020, pp. 173--174.

\bibitem{wang2016deepburning}
Y.~Wang, J.~Xu, Y.~Han, H.~Li, and X.~Li, ``Deepburning: Automatic generation of fpga-based learning accelerators for the neural network family,'' in \emph{Proceedings of the 53rd Annual Design Automation Conference}, 2016, pp. 1--6.

\bibitem{zhang2018dnnbuilder}
X.~Zhang, J.~Wang, C.~Zhu, Y.~Lin, J.~Xiong, W.-m. Hwu, and D.~Chen, ``Dnnbuilder: An automated tool for building high-performance dnn hardware accelerators for fpgas,'' in \emph{2018 IEEE/ACM International Conference on Computer-Aided Design (ICCAD)}.\hskip 1em plus 0.5em minus 0.4em\relax IEEE, 2018, pp. 1--8.

\bibitem{wei2017automated}
X.~Wei, C.~H. Yu, P.~Zhang, Y.~Chen, Y.~Wang, H.~Hu, Y.~Liang, and J.~Cong, ``Automated systolic array architecture synthesis for high throughput cnn inference on fpgas,'' in \emph{Proceedings of the 54th Annual Design Automation Conference 2017}, 2017, pp. 1--6.

\bibitem{samajdar2018scale}
A.~Samajdar, Y.~Zhu, P.~Whatmough, M.~Mattina, and T.~Krishna, ``Scale-sim: Systolic cnn accelerator simulator,'' \emph{arXiv preprint arXiv:1811.02883}, 2018.

\bibitem{luo2023deepburning}
E.~Luo, H.~Huang, C.~Liu, G.~Li, B.~Yang, Y.~Wang, H.~Li, and X.~Li, ``Deepburning-mixq: An open source mixed-precision neural network accelerator design framework for fpgas,'' in \emph{2023 IEEE/ACM International Conference on Computer Aided Design (ICCAD)}.\hskip 1em plus 0.5em minus 0.4em\relax IEEE, 2023, pp. 1--9.

\bibitem{montgomerie2023pass}
A.~Montgomerie-Corcoran, Z.~Yu, J.~Cheng, and C.-S. Bouganis, ``Pass: Exploiting post-activation sparsity in streaming architectures for cnn acceleration,'' in \emph{2023 33rd International Conference on Field-Programmable Logic and Applications (FPL)}.\hskip 1em plus 0.5em minus 0.4em\relax IEEE, 2023, pp. 288--293.

\bibitem{petrica2020memory}
L.~Petrica, T.~Alonso, M.~Kroes, N.~Fraser, S.~Cotofana, and M.~Blott, ``Memory-efficient dataflow inference for deep cnns on fpga,'' in \emph{2020 International Conference on Field-Programmable Technology (ICFPT)}.\hskip 1em plus 0.5em minus 0.4em\relax IEEE, 2020, pp. 48--55.

\bibitem{alonso2021elastic}
T.~Alonso, L.~Petrica, M.~Ruiz, J.~Petri-Koenig, Y.~Umuroglu, I.~Stamelos, E.~Koromilas, M.~Blott, and K.~Vissers, ``Elastic-df: Scaling performance of dnn inference in fpga clouds through automatic partitioning,'' \emph{ACM Transactions on Reconfigurable Technology and Systems (TRETS)}, vol.~15, no.~2, pp. 1--34, 2021.

\bibitem{Ibrahim2023extending}
M.~Ibrahim, Z.~Zhao, M.~Hall, and V.~Betz, ``Extending data flow architectures for convolutional neural networks to multiple fpgas,'' in \emph{2023 International Conference on Field-Programmable Technology (ICFPT)}.\hskip 1em plus 0.5em minus 0.4em\relax IEEE, 2023, pp. 126--135.

\bibitem{montgomerie2023satay}
A.~Montgomerie-Corcoran, P.~Toupas, Z.~Yu, and C.-S. Bouganis, ``Satay: a streaming architecture toolflow for accelerating yolo models on fpga devices,'' \emph{arXiv preprint arXiv:2309.01587}, 2023.

\bibitem{yu2023autows}
Z.~Yu and C.-S. Bouganis, ``Autows: Automate weights streaming in layer-wise pipelined dnn accelerators,'' 2023.

\bibitem{simonyan2014very}
K.~Simonyan and A.~Zisserman, ``Very deep convolutional networks for large-scale image recognition,'' \emph{arXiv preprint arXiv:1409.1556}, 2014.

\bibitem{szegedy2015going}
C.~Szegedy, W.~Liu, Y.~Jia, P.~Sermanet, S.~Reed, D.~Anguelov, D.~Erhan, V.~Vanhoucke, and A.~Rabinovich, ``Going deeper with convolutions,'' in \emph{Proceedings of the IEEE conference on computer vision and pattern recognition}, 2015, pp. 1--9.

\bibitem{he2016deep}
K.~He, X.~Zhang, S.~Ren, and J.~Sun, ``Deep residual learning for image recognition,'' in \emph{Proceedings of the IEEE conference on computer vision and pattern recognition}, 2016, pp. 770--778.

\bibitem{howard2017mobilenets}
A.~G. Howard, M.~Zhu, B.~Chen, D.~Kalenichenko, W.~Wang, T.~Weyand, M.~Andreetto, and H.~Adam, ``Mobilenets: Efficient convolutional neural networks for mobile vision applications,'' \emph{arXiv preprint arXiv:1704.04861}, 2017.

\bibitem{lecun1998mnist}
Y.~LeCun, ``The mnist database of handwritten digits,'' \emph{http://yann. lecun. com/exdb/mnist/}, 1998.

\bibitem{krizhevsky2009learning}
A.~Krizhevsky, G.~Hinton \emph{et~al.}, ``Learning multiple layers of features from tiny images,'' 2009.

\bibitem{deng2009imagenet}
J.~Deng, W.~Dong, R.~Socher, L.-J. Li, K.~Li, and L.~Fei-Fei, ``Imagenet: A large-scale hierarchical image database,'' in \emph{2009 IEEE conference on computer vision and pattern recognition}.\hskip 1em plus 0.5em minus 0.4em\relax Ieee, 2009, pp. 248--255.

\bibitem{nakahara2018lightweight}
H.~Nakahara, H.~Yonekawa, T.~Fujii, and S.~Sato, ``A lightweight yolov2: A binarized cnn with a parallel support vector regression for an fpga,'' in \emph{Proceedings of the 2018 ACM/SIGDA International Symposium on field-programmable gate arrays}, 2018, pp. 31--40.

\bibitem{anupreetham2023high}
A.~Anupreetham, M.~Ibrahim, M.~Hall, A.~Boutros, A.~Kuzhively, A.~Mohanty, E.~Nurvitadhi, V.~Betz, Y.~Cao, and J.-s. Seo, ``High throughput fpga-based object detection via algorithm-hardware co-design,'' \emph{ACM Transactions on Reconfigurable Technology and Systems}, 2023.

\bibitem{toupas2023fpgahart}
P.~Toupas, C.-S. Bouganis, and D.~Tzovaras, ``fpgahart: A toolflow for throughput-oriented acceleration of 3d cnns for har onto fpgas,'' \emph{arXiv preprint arXiv:2305.19896}, 2023.

\bibitem{toupas2023harflow3d}
P.~Toupas, A.~Montgomerie-Corcoran, C.-S. Bouganis, and D.~Tzovaras, ``Harflow3d: A latency-oriented 3d-cnn accelerator toolflow for har on fpga devices,'' in \emph{2023 IEEE 31st Annual International Symposium on Field-Programmable Custom Computing Machines (FCCM)}, 2023, pp. 144--154.

\bibitem{montgomerie2022samo}
A.~Montgomerie-Corcoran, Z.~Yu, and C.-S. Bouganis, ``Samo: Optimised mapping of convolutional neural networks to streaming architectures,'' in \emph{2022 32nd International Conference on Field-Programmable Logic and Applications (FPL)}.\hskip 1em plus 0.5em minus 0.4em\relax IEEE, 2022, pp. 418--424.

\bibitem{yolov8_ultralytics}
\BIBentryALTinterwordspacing
G.~Jocher, A.~Chaurasia, and J.~Qiu, ``Ultralytics yolov8,'' 2023. [Online]. Available: \url{https://github.com/ultralytics/ultralytics}
\BIBentrySTDinterwordspacing

\bibitem{cicek2016unet3d}
{\"O}.~{\c{C}}i{\c{c}}ek, A.~Abdulkadir, S.~S. Lienkamp, T.~Brox, and O.~Ronneberger, ``3d u-net: Learning dense volumetric segmentation from sparse annotation,'' in \emph{Medical Image Computing and Computer-Assisted Intervention -- MICCAI 2016}, S.~Ourselin, L.~Joskowicz, M.~R. Sabuncu, G.~Unal, and W.~Wells, Eds.\hskip 1em plus 0.5em minus 0.4em\relax Cham: Springer International Publishing, 2016, pp. 424--432.

\bibitem{lin2014microsoft}
T.-Y. Lin, M.~Maire, S.~Belongie, J.~Hays, P.~Perona, D.~Ramanan, P.~Doll{\'a}r, and C.~L. Zitnick, ``Microsoft coco: Common objects in context,'' in \emph{Computer Vision--ECCV 2014: 13th European Conference, Zurich, Switzerland, September 6-12, 2014, Proceedings, Part V 13}.\hskip 1em plus 0.5em minus 0.4em\relax Springer, 2014, pp. 740--755.

\bibitem{brostow2009semantic}
G.~J. Brostow, J.~Fauqueur, and R.~Cipolla, ``Semantic object classes in video: A high-definition ground truth database,'' \emph{Pattern Recognition Letters}, vol.~30, no.~2, pp. 88--97, 2009.

\bibitem{menze2014multimodal}
B.~H. Menze, A.~Jakab, S.~Bauer, J.~Kalpathy-Cramer, K.~Farahani, J.~Kirby, Y.~Burren, N.~Porz, J.~Slotboom, R.~Wiest \emph{et~al.}, ``The multimodal brain tumor image segmentation benchmark (brats),'' \emph{IEEE transactions on medical imaging}, vol.~34, no.~10, pp. 1993--2024, 2014.

\bibitem{soomro2012ucf101}
K.~Soomro, A.~R. Zamir, and M.~Shah, ``Ucf101: A dataset of 101 human actions classes from videos in the wild,'' \emph{arXiv preprint arXiv:1212.0402}, 2012.

\bibitem{basalama2023flexcnn}
\BIBentryALTinterwordspacing
S.~Basalama, A.~Sohrabizadeh, J.~Wang, L.~Guo, and J.~Cong, ``Flexcnn: An end-to-end framework for composing cnn accelerators on fpga,'' vol.~16, no.~2, mar 2023. [Online]. Available: \url{https://doi.org/10.1145/3570928}
\BIBentrySTDinterwordspacing

\bibitem{liu2019towards}
S.~Liu and W.~Luk, ``Towards an efficient accelerator for dnn-based remote sensing image segmentation on fpgas,'' in \emph{2019 29th International Conference on Field Programmable Logic and Applications (FPL)}, 2019, pp. 187--193.

\end{thebibliography}
